\documentclass[12pt,onecolumn,draftclsnofoot,journal]{IEEEtran}
\usepackage{graphicx}
\usepackage{float}
\usepackage{epstopdf}
\usepackage[cmex10]{amsmath}
\usepackage{array}
\usepackage{cite}
\usepackage{amssymb}
\usepackage{amsfonts}
\usepackage{amsmath}
\usepackage{stackrel}
\usepackage{booktabs,multirow}
\usepackage{arydshln}
\usepackage{slashbox}
\usepackage{amsthm}
\usepackage{listings}
\usepackage{algorithm}
\usepackage{algorithmicx}
\usepackage{algpseudocode}
\usepackage{tablefootnote}
\usepackage{threeparttable}

\newtheorem{lem}{Lemma}
\newtheorem{rem}{Remark}
\newtheorem{theo}{Theorem}

\newtheorem{cor}{Corollary}
\newtheorem{pro}{Proposition}
\newtheorem{ex}{Example}
\newfloat{routine}{htbp}{loa}
\floatname{routine}{Routine} 

\makeatletter
\newcommand{\algmargin}{\the\ALG@thistlm}
\makeatother
\newlength{\forwidth}
\algdef{SE}[parFOR]{parFor}{EndparFor}[1]
  {\parbox[t]{\dimexpr\linewidth-\algmargin}{
     \hangindent\forwidth\strut\algorithmicfor\ #1\ \algorithmicdo\strut}}{\algorithmicend\ \algorithmicfor}
\algnewcommand{\parState}[1]{\State
  \parbox[t]{\dimexpr\linewidth-\algmargin}{\strut #1\strut}}

\newlength{\ifwidth}
\algdef{SE}[parIF]{parIf}{EndparIf}[1]
  {\parbox[t]{\dimexpr\linewidth-\algmargin}{
     \hangindent\ifwidth\strut\algorithmicif\ #1\ \algorithmicdo\strut}}{\algorithmicend\ \algorithmicif}

\begin{document}

\title{Characterization and Efficient Exhaustive Search Algorithm for Elementary Trapping Sets of Irregular LDPC Codes}
\author{Yoones Hashemi,\IEEEmembership{ Student Member, IEEE}, and Amir H. Banihashemi,\IEEEmembership{ Senior Member, IEEE}}
\maketitle
\IEEEpeerreviewmaketitle

\begin{abstract}
In this paper, we propose a characterization of elementary trapping sets (ETSs) for irregular low-density parity-check (LDPC) codes. These sets 
are known to be the main culprits in the error floor region of such codes. The characterization of ETSs for irregular codes has been known to be a 
challenging problem due to the large variety of non-isomorphic ETS structures that can exist within the Tanner graph of these codes.
This is a direct consequence of the variety of the degrees of the variable nodes that can participate 
in such structures. The proposed characterization is based on a hierarchical graphical representation of
ETSs, starting from simple cycles of the graph, or from single variable nodes, and involves three simple expansion 
techniques: degree-one tree ($dot$), $path$ and $lollipop$, thus, the terminology {\em dpl characterization}. 
A similar dpl characterization was proposed in an earlier work by the authors for the leafless ETSs (LETSs) of variable-regular 
LDPC codes. The present paper generalizes the prior work to codes with a variety of variable node degrees and to 
ETSs that are not leafless. 
The proposed dpl characterization corresponds to an efficient search algorithm that, 
for a given irregular LDPC code, can find all the instances of $(a,b)$ ETSs with size $a$ and with the number of unsatisfied check nodes $b$ within
any range of interest $a \leq a_{max}$ and $b \leq b_{max}$, exhaustively.
Although, (brute force) exhaustive search algorithms for ETSs of irregular LDPC codes exist, to the best of our knowledge, 
the proposed search algorithm is the first of its kind, in that, it is devised based on a characterization of ETSs that makes the search process efficient.
Extensive simulation results are presented to show the versatility of the search algorithm, and to demonstrate that, compared to the literature, 
significant improvement in search speed can be obtained. 
\end{abstract}

\section{introduction}

 It is well-known that error-floor performance of low-density parity-check (LDPC) codes is related to the presence of certain problematic graphical structures in the Tanner graph of the code, commonly referred to as \textit{trapping sets} (TS)~\cite{richardson}.
Among TSs, the most harmful ones are known to be the \textit{elementary trapping sets (ETSs)}~\cite{milenkovic2007asymptotic},\cite{zhang2009toward},\cite{laendner2009},
whose induced subgraphs contain only degree-1 and degree-2 check nodes.  In particular, the \textit{leafless} ETSs (LETSs), in which each variable node is connected to at least two even-degree (satisfied) check nodes, 
are recognized as the main culprit for variable-regular LDPC codes \cite{hashemireg}. 
For a given LDPC code, the knowledge of trapping sets is important. Such knowledge can be used, for example, to estimate the error floor~\cite{richardson},\cite{sina1}, to devise decoding algorithms with low error floor~\cite{zhang2009toward},\cite{kyung2012finding}, or 
to design codes with low error floor~\cite{Ivkovic},~\cite{Asvadi},~\cite{ABA-2012},\cite{KAB-2012},\cite{nguyen}. There are numerous works on the characterization and search algorithms for trapping sets    \cite{Wang}, \cite{vasic2009trapping},\cite{wang2009finding}, \cite{abu2010trapping}, \cite{Rosnes},\cite{Dolecek}, \cite{Diao},\cite{laendner2010characterization},  \cite{dolecek2010analysis}, \cite{schlegel2010dynamics},\cite{zhang2011efficient}, \cite{mehdi2012}, \cite{nguyen},\cite{kyung2012finding}, \cite{Deka13}, \cite{mehdi2014}, \cite{yoones2015}, \cite{hashemireg}. Most of these works, however, are concerned with variable-regular LDPC codes. 

There are very few works on the trapping sets of irregular LDPC codes~\cite{abu2010trapping},\cite{mehdi2012},\cite{kyung2012finding},\cite{Fals}. This is despite the fact that these codes are popular in many applications due to their superior performance over the regular codes in the waterfall region~\cite{richardson2}.
In fact, irregular LDPC codes have been already adopted in a number of standards~\cite{802.16},\cite{802.11},~\cite{DVB-S2},~\cite{DVB-T2},~\cite{802.22}. One main reason for the lack of results on trapping sets of irregular codes is the variety of variable node degrees that makes the identification and characterization of trapping sets of different classes a seemingly impossible task~\cite{mehdi2012}. In the following, we review the main existing literature related to trapping sets of irregular codes: In \cite{abu2010trapping}, using the modified impulse algorithm, the authors devised a technique
to find a non-exhaustive list of trapping sets of a given irregular LDPC code. In \cite{mehdi2012}, by examining the relationships between cycles and trapping sets, 
Karimi and Banihashemi proposed an efficient search algorithm to find the dominant ETSs of a given irregular LDPC code. Although the proposed algorithm 
in \cite{mehdi2012} can find ETSs with sufficiently large size, it provides no guarantee that the obtained list of ETSs is exhaustive.
Using the branch-\&-bound principle, Kyung and Wang \cite{kyung2012finding} proposed an exhaustive search algorithm to enumerate the fully absorbing sets
(FASs) of short irregular LDPC codes. The proposed algorithm, however, becomes quickly infeasible to use as the block length, $n$, and the size of the FASs, $a$, are increased. In
general, the reach of the algorithm is limited to $n < 1000$ and $a < 7$, or $n < 1000$ and $a < 14$, if the number of unsatisfied check nodes is limited to $b < 3$. 
Very recently, Falsafain and Mousavi \cite{Fals} proposed a branch-\&-bound algorithm to find the ETSs of irregular LDPC codes. Although, the proposed $0-1$ integer linear programming (ILP) formulation in \cite{Fals} provides a tighter linear programming relaxation in comparison with the one in \cite{kyung2012finding}, the exhaustive search algorithm of \cite{Fals} is still only applicable to short-to-moderate length LDPC codes.
To the best of our knowledge, the algorithms proposed in \cite{kyung2012finding} and \cite{Fals} are the only exhaustive search algorithms of error-prone structures in irregular LDPC codes.  

In \cite{mehdi2014}, \cite{yoones2015}, \cite{hashemi2015corrections}, the authors studied the graphical structure of LETSs in variable-regular LDPC codes and 
demonstrated that all the non-isomorphic structures of LETSs are layered super sets (LSS) of some basic structures. 
This characterization corresponds to a search algorithm that can find all the instances of LETSs in a guaranteed fashion.
Although, the search algorithm itself is simple, one may need to enumerate basic structures with large
size in the Tanner graph of the code, as the input to the search algorithm. The multiplicity of such structures increases rapidly with the size and thus the enumeration, storage and processing of these structures pose a practical hurdle in implementing the search algorithm. To overcome this problem, in an earlier work~\cite{hashemireg}, we proposed a novel hierarchical graph-based expansion approach to characterize LETSs of variable-regular LDPC codes.
The proposed $dpl$ characterization is based on three basic expansion techniques, dubbed, $dot$, $path$ and $lollipop$. Each LETS structure ${\cal S}$ is characterized as a sequence of embedded LETS structures that starts from a simple cycle, and grows in each step by using one of the three expansions, until it reaches ${\cal S}$. 
The new characterization, allowed us to devise search algorithms that are provably efficient in finding all the instances 
of $(a,b)$ LETS structures with $a \leq a_{max}$ and $b \leq b_{max}$, for any choice of $a_{max}$ and $b_{max}$, in a guaranteed fashion.
To the best of our knowledge, the proposed algorithm in \cite{hashemireg} is the most efficient exhaustive search algorithm available for finding LETSs of variable-regular LDPC codes.
It is also the most general, in that, it is applicable to codes with any variable degree, girth, rate and block length. 

In this paper, we generalize the approach proposed in \cite{hashemireg} to irregular LDPC codes.
We develop a framework to use the dpl characterization of LETS structures in variable-regular graphs to characterize the LETS structures of irregular graphs (in a given range of $a$ and $b$ values, exhaustively). This characterization corresponds to an exhaustive search algorithm. The complexity and memory requirements of the search, however, grows rapidly with the increase in the range of $a$ and $b$ values and the variety of variable node degrees, 
particularly for irregular codes with degree-$2$ variable nodes. For such scenarios, we propose an alternate approach to characterize/search LETSs of irregular codes.
The new approach, which is still based on the three expansion techniques $dot$, $path$ and $lollipop$, uses a sequence of recursively derived upper bounds $b^{a'}_{max}$ on the maximum number of unsatisfied check nodes $b$ 
that can possibly appear in a substructure of size $a'$ of a LETS structure within the interest range of $a \leq a_{max}$ and $b \leq b_{max}$.   
As another contribution, we extend our characterization/exhaustive search to ETSs that are not leafless. This together with our results on LETSs, provide an efficient exhaustive search algorithm for all ETSs (leafless and otherwise) of irregular LDPC 
codes. We then apply the search algorithm to a large number of irregular LDPC codes with a variety of degree distributions, girths, rates and block lengths, to demonstrate the strength and the versatility of the proposed scheme.
We note that compared to the IP-based search methods of \cite{kyung2012finding} and \cite{Fals}, the proposed algorithm is more efficient and has a much wider reach for finding problematic structures of larger size $a$ within 
codes of larger block length $n$. The main reason for this superiority is that, unlike the brute force algorithms of \cite{kyung2012finding} and \cite{Fals}, the proposed scheme uses carefully devised embedded sequences of structures 
each starting from a simple cycle, and then expanding step by step to larger structures using one of the three simple expansion techniques.  

The remainder of this paper is organized as follows. Basic definitions and notations are provided in Section \ref{sec:pre}.  In Section \ref{sec:regular}, the dpl-based characterization/search for variable-regular graphs is revisited. In Section \ref{sec:char/search}, a novel approach is proposed to extend the dpl characterization/search to irregular graphs. In Section \ref{sec:search}, the shortcomings of the approach proposed in Section~\ref{sec:char/search} are discussed, and 
an alternate characterization/search is developed to address those shortcomings. Characterization of ETSs that are not leafless, and an efficient exhaustive search algorithm to find them
are presented in Section \ref{sec:ETS}. 
Finally, numerical results are provided in Section \ref{sec:numerical}, followed by concluding remarks in Section \ref{sec:conclude}.

\section{Preliminaries }

\label{sec:pre}

Consider an undirected graph $G=(F, E)$, where the two sets $F=\{f_1,\dots,f_k\}$ and $E=\{e_1,\dots,e_m\}$, are the sets of \textit{nodes} and \textit{edges} of $G$,
respectively. 
We say that an edge $e$ is \textit{incident} to a node $f$ if $e$ is connected to $f$.  If there exists an edge $e_k$ which is incident to 
two distinct nodes $f_i$ and $f_j$, we represent $e_k$ by $f_i f_j$ or $f_j f_i$.   
The degree of a node $f$ is denoted by $d_f$, and is defined as the number of edges incident to $f$. 
The \textit{maximum degree} and the \textit{minimum degree} of a graph $G$,  denoted by $\Delta (G)$ and $\delta (G)$, respectively, are defined to be the maximum and minimum 
degree of its nodes, respectively. 

Given an undirected graph $G=(F,E)$,
 a \textit{walk} between two nodes $f_1$ and $f_{k+1}$ is a sequence of nodes and edges
 $f_1$, $e_1$, $f_2$, $e_2$, $\dots$, $f_k$, $e_k$, $f_{k+1}$, where $e_i=f_i f_{i+1}$, $\forall i \in [1,k]$. 
In this definition, the nodes $f_1,f_2,\dots,f_{k+1}$ are not necessarily distinct. The same applies to the edges $e_1,e_2,\dots,e_k$.
 A {\em path} is a walk  with no repeated nodes or edges, except the first and the last nodes that can be the same. If the first and the last nodes are distinct, we call the path an {\em open path}.
Otherwise, we call it a {\em closed path} or a {\em cycle}. The \textit{length} of a walk, a  path, or a cycle is the number of its edges. 
A {\em lollipop walk} is a walk $f_1$, $e_1$, $f_2$, $e_2$, $\dots$, $f_k$, $e_k$, $f_{k+1}$, such that all the edges and all the nodes are distinct, except that $f_{k+1}=f_m$, for some $m \in (1,k)$.
A \textit{chord} of a cycle is an edge which is not part of the cycle but is incident to two distinct nodes in the cycle.
A \textit{simple cycle} or a {\em chordless cycle} is a cycle which does not have any chord. 
Throughout this paper, we use the notation $s_k$ for a simple cycle of length $k$. The length of the shortest cycle in a graph is called \textit{girth}, and is denoted by $g$.

A graph is called {\em connected} when there is a {\em path} between every pair of nodes. A \textit{tree} is a connected graph that contains no cycles. 
A \textit{rooted tree} is a tree in which one specific node is assigned as the \textit{root}. The \textit{depth of a node} in a rooted tree is the length of the
path from the node to the root. The \textit{depth of a tree} is the maximum depth of any node in the tree. 
\textit{Depth-one tree} is a tree with depth one. 
A node $f$ is called \textit{leaf} if $d_f=1$. Although this terminology is commonly used for trees, in this paper, we use it for a general graph that may contain cycles. 
A \textit{leafless graph} is a connected graph $G$ with $\delta (G) \geq 2$.

The graphs $G_1 =(F_1,E_1)$ and $G_2 =(F_2,E_2)$ are
\textit{isomorphic} if there is a bijection $p : F_1 \rightarrow F_2$ such that
nodes $f_1, f_2 \in F_1$ are joined by an edge if and only if $p(f_1)$
and $p(f_2)$ are joined by an edge. Otherwise, the graphs are \textit{non-isomorphic}.

Any $m \times n$ parity check matrix $H$ of a binary LDPC code $\mathcal{C}$ can be represented by its bipartite Tanner graph $G=(V \cup C, E)$, 
where $V=\{ v_1,v_2,\dots,v_n \}$ is the set of variable nodes and $C=\{ c_1,c_2,\dots,c_m \}$ is the set of check nodes. An edge $e=v_i c_j$ in $E$ corresponds to a $1$ in the $(j,i)$ entry of matrix $H$.
A Tanner graph is called {\em variable-regular} with variable degree $d_{\mathrm{v}}$ if $d_{v_i} = d_{\mathrm{v}}$, $\forall~{v}_{i} \in V$. 
A Tanner graph is called {\em irregular} if it has multiple variable or check node degrees. The node degrees for an irregular LDPC code are often described by the code's variable and check node degree distributions, $\lambda(x)=\sum\limits_{i=d_{v_{min}}}^{d_{v_{max}}} \lambda_i x^{i-1}$ and $\rho(x)=\sum\limits_{i=d_{c_{min}}}^{d_{c_{max}}} \rho_i x^{i-1}$, respectively, 
where $\lambda_i$ and $\rho_i$ are the fractions of edges in the Tanner graph that are incident to degree-$i$ variable and degree-$i$ check nodes, respectively. 
The terms $d_{v_{max}}$ and $d_{v_{min}}$ ($d_{c_{max}}$ and $d_{c_{min}}$) are the maximum and minimum degrees of variable nodes (check nodes), respectively.
The length of cycles in a Tanner graph can only be an even number. We study the Tanner graphs that are free of 4-cycles ($g > 4$).

For a subset $\mathcal{S}$ of $V$, the subset $\Gamma{(\mathcal{S})}$ of $C$ denotes the set of neighbors of $\mathcal{S}$ in $G$.
The \textit{induced subgraph} of $\mathcal{S}$ in $G$, denoted by $G(\mathcal{S})$, is the graph with the set of nodes $\mathcal{S} \cup \Gamma{(\mathcal{S})}$ and 
the set of edges $\{f_i f_j \in E : f_i \in \mathcal{S}, f_j \in \Gamma{(\mathcal{S})}\}$. The set of check nodes with odd and even degrees in $G(\mathcal{S})$ are denoted by $\Gamma_{o}{(\mathcal{S})}$ and $\Gamma_{e}{(\mathcal{S})}$, respectively. In this paper, the terms \textit{unsatisfied check nodes} and \textit{satisfied check nodes} are used to refer to the check nodes in
$\Gamma_{o}{(\mathcal{S})}$ and $\Gamma_{e}{(\mathcal{S})}$, respectively. 
The \textit{size} of an induced subgraph $G(\mathcal{S})$ is defined to be the number of its variable nodes.  
We assume that an induced subgraph is connected. Disconnected subgraphs can be considered as the union of connected ones. 
All the induced subgraphs with the same size $a$, 
and the same number of unsatisfied check nodes $b$, are considered to belong to the same \textit{$(a,b)$ class}.

Given a Tanner graph G,
a set $\mathcal{S}\subset V$ is called an \textit{(a,b) trapping set (TS)} if $|\mathcal{S}| = a$ and $|\Gamma_{o}{(\mathcal{S})}| = b$.
Alternatively, $\mathcal{S}$ is said to belong to the {\em class of (a,b) TSs}. Parameter $a$ is referred to as the {\em size} of the TS.
In the rest of the paper, depending on the context, the term ``trapping set'' may be used to refer to the set of variable nodes $\mathcal{S}$, or to the induced subgraph $G(\mathcal{S})$ of $\mathcal{S}$ in the Tanner graph $G$. Similarly, we may use ${\cal S}$ to mean $G(\mathcal{S})$.
 An \textit{elementary trapping set (ETS)} is a trapping set for which all the check nodes in $G(\mathcal{S})$ have degree 1 or 2.
 A set $\mathcal{S}\subset V$ is called an \textit{(a,b) absorbing set (AS)} if $\mathcal{S}$ is an $(a,b)$ trapping set and all
 the variable nodes in $\mathcal{S}$  are connected to more nodes in $\Gamma_{e}{(\mathcal{S})}$ than in $\Gamma_{o}{(\mathcal{S})}$.
 An {\em elementary absorbing set (EAS)} $\mathcal{S}$ is an absorbing set for which all the check nodes in $G(\mathcal{S})$ have degree 1 or 2.
 A \textit{fully absorbing set (FAS)} $\mathcal{S}\subset V$ is an absorbing set for which all the nodes in $V \backslash \mathcal{S}$ have strictly more neighbors in $C \backslash \Gamma_{o}{(\mathcal{S})}$ than in $\Gamma_{o}{(\mathcal{S})}$.
A set $\mathcal{S}\subset V$ is called an \textit{(a,b) fully elementary absorbing set (FEAS)} if $\mathcal{S}$ is an $(a,b)$
EAS and if all the nodes in $V\backslash \mathcal{S}$ have strictly more neighbors  in $C\backslash \Gamma_{o}{(\mathcal{S})}$ than in $\Gamma_{o}{(\mathcal{S})}$.

Elementary trapping sets are the subject of this paper. To simplify the representation of ETSs, similar to ~\cite{mehdi2014}, \cite{yoones2015}, \cite{hashemireg}, we use an 
alternate graph representation of ETSs, called \textit{normal graph}.
The normal graph of an ETS $\mathcal{S}$ is obtained from $G(\mathcal{S})$ by removing all the check nodes of degree one and their incident edges, and by replacing all the 
degree-2 check nodes and their two incident edges by a single edge. It is easy to see that there is a one-to-one correspondence between the Tanner graph
$G(\mathcal{S})$ and the normal graph of $\mathcal{S}$ for variable-regular LDPC codes. 

\begin{lem}
\label{lem:eb}
Consider the normal graph of an $(a,b)$ ETS structure of a variable-regular Tanner graph with variable degree $d_{\mathrm{v}}$.
The number of nodes and edges of this normal graph are $a$ and $(a d_{\mathrm{v}} - b)/2$, respectively. We thus have $b=a d_{\mathrm{v}} - 2 e$,
where $e$ is the number of edges of the normal graph. 
\label{lem1}
\end{lem}

 We call a set $\mathcal{S}\subset V$ an \textit{(a,b) leafless ETS (LETS)} if $\mathcal{S}$ is an $(a,b)$
 ETS and if the normal graph of $\mathcal{S}$ is leafless. 
\begin{ex}
 Fig. \ref{fig:leafless}(a) represents a LETS in the $(4,2)$ class in a variable-regular Tanner graph with $d_{\mathrm{v}}=3$ and its leafless normal graph.
 Fig. \ref{fig:leafless}(b) shows an ETS in the $(4,4)$ class and its normal graph with a leaf. 
(Symbols \scalebox{0.7}{$\square$} and \scalebox{0.7}{$\blacksquare$} are used to represent satisfied and unsatisfied check nodes in the induced subgraphs, respectively, and the symbol 
\scalebox{1.2}{$\circ$} is used to represent variable nodes in both the induced subgraphs and normal graphs.) 
\begin{figure}[] 
\centering
\includegraphics [width=0.35\textwidth]{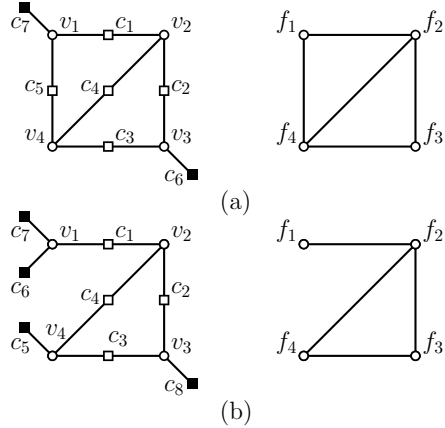}
\caption{(a) A LETS in the $(4,2)$ class and its leafless normal graph, (b) An ETS in the $(4,4)$ class and its normal graph which has a leaf $(f_1)$.}
\label{fig:leafless}
\end{figure}
\end{ex}

Unlike the variable-regular Tanner graphs, there is not a one-to-one correspondence between an ETS and its normal graph for irregular graphs.
In other words, for an irregular Tanner graph, the number of edges in the normal graph representation of an ETS is not uniquely mapped to the number of unsatisfied check nodes in the ETS. 
To have a one-to-one correspondence, in addition to the normal graph, the extra information about the degrees of variable nodes involved in the ETS is also required.
For this, we introduce a new graphical representation of an ETS, which we call \textit{quasi-normal} representation.
The \textit{quasi-normal graph} of an ETS $\mathcal{S}$ is obtained from $G(\mathcal{S})$ by  replacing all the 
check nodes (degree-one or two) and their incident edges by a single edge. In this representation, the edges that are connected to only one node (singly-connected edges)
are responsible for preserving the degree of variable nodes. 
It is easy to see that there is a one-to-one correspondence between $G(\mathcal{S})$ and the quasi-normal graph of $\mathcal{S}$ for any regular or irregular LDPC code.
In the following, we still continue to use the normal graph representation for irregular graphs. 
Such a representation can be considered as the image or projection of quasi-normal graphs into the space of normal graphs, where such a projection involves dropping all the singly-connected edges.
In general, in an irregular Tanner graph, multiple ETS structures with different quasi-normal graphs may have the same normal graph representation.
We also continue to use the same definition of LETS for irregular graphs, i.e., an ETS $\mathcal{S}$ is LETS if the normal graph of $\mathcal{S}$ is leafless.

\begin{ex}
Fig. \ref{fig:notnormal} shows the induced subgraphs of four LETS structures in an irregular Tanner graph with variable node degrees 3 and 4. The figure also includes the corresponding quasi-normal graphs and the normal graph representation 
of the four structures. It can be seen that all four non-isomorphic LETS structures have the same normal graph.
\begin{figure}[] 
\centering
\includegraphics [width=0.5\textwidth]{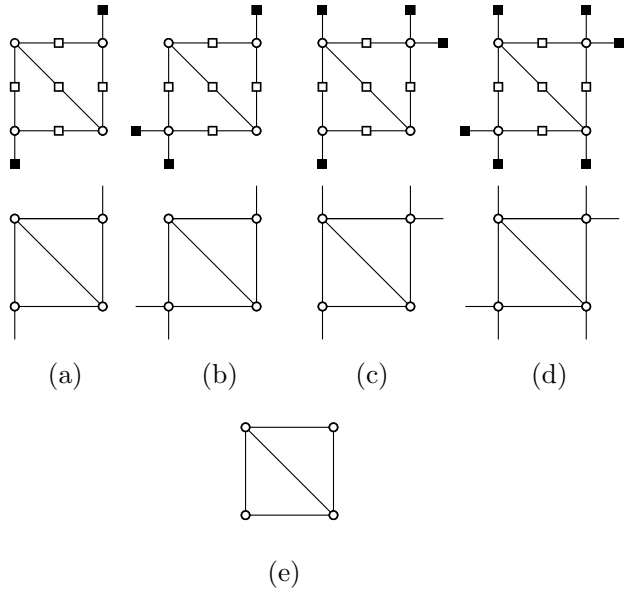}
\caption{Tanner graph and quasi-normal graph representations of LETSs in an irregular Tanner graph with variable degrees 3 and 4 in (a) $(4,2)$, (b) $(4,3)$, (c) $(4,4)$, (d) $(4,6)$ classes and (c) their normal graph.}
\label{fig:notnormal}
\end{figure}
\end{ex}

To the best of our knowledge, almost all the structures reported in the literature as error-prone structures of irregular LDPC codes are ETSs, with a large majority being LETS structures. LETSs have thus been the subject of many studies including~\cite{Tse2010},\cite{mehdi2012},\cite{butler2014},\cite{mehdi2014},\cite{yoones2015},\cite{hashemireg}. In this paper, we first focus on LETS structures of irregular codes in Sections \ref{sec:char/search} and \ref{sec:search}. We then study ETS structures which are not leafless in Section \ref{sec:ETS}.

In the following, when we are concerned with the characterization of ETS structures, we use the quasi-normal/normal graph representation.
On the other hand, when we discuss search algorithms, since the search is performed on a Tanner graph, we are concerned with the bipartite graph representation of an ETS structure. Nevertheless, for consistency and to prevent confusion, even in the context of search algorithms, we still use terminologies corresponding to normal graphs.
For example, we use ``all the instances of $s_k$'' to mean ``all the instances of the structure whose normal graph is $s_k$."

\section{Dpl Characterization/Search for Variable-Regular LDPC Codes}

\label{sec:regular}

In an earlier work~\cite{hashemireg}, we developed a characterization/search algorithm for LETS structures of variable-regular LDPC codes in the space of normal graphs. 
In the dpl characterization, each LETS structure is identified with an embedded sequence of LETS structures
that starts from a simple cycle and expands, at each step, to a larger LETS structure using one of the three expansions, $dot$, $path$ or $lollipop$, until it reaches the LETS structure of interest. In
this characterization, the simple cycle is called a {\em prime} structure with respect to dpl expansions.  
In the following, we briefly explain the three expansions.

\begin{figure}[] 
\centering
\includegraphics [width=0.5\textwidth]{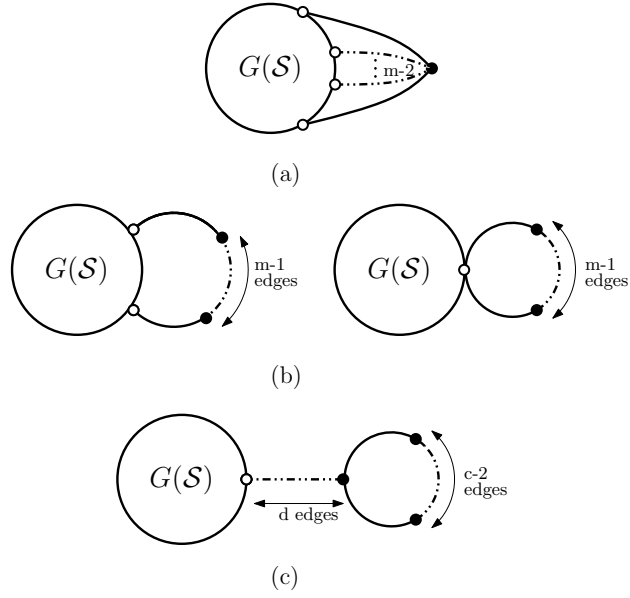}
\caption{Expansion of the LETS structure $\mathcal{S}$ with (a) a depth-one tree with $m$ edges, $dot_m$, (b) an open and closed $path$ of length $m+1, pa_m^o, pa_m^c$, respectively, (c) a lollipop walk of length $m+1=d+c$, $lo^c_m$.}
\label{fig:expans}
\end{figure}

Consider an $(a,b)$ LETS structure ${\cal S}$ of a variable-regular Tanner graph with $g \geq 6$ and variable degree $d_{\mathrm{v}}$. 
Figs. \ref{fig:expans}(a)-(c) show the three expansions applied to the induced subgraph of ${\cal S}$. In these figures, the symbol \scalebox{1.2}{$\circ$} 
is used to represent the common node(s) between $\mathcal{S}$ and the expansion graph, and the symbol \scalebox{1.2}{$\bullet$} is used to represent the other nodes of the expansion graph.

In Fig. \ref{fig:expans}(a), the expansion using a depth-one tree is shown. 
The notation $dot_m$ is used for a {\em depth-one tree (dot)} expansion with  $m$ edges.
For the $dot_m$ expansion to result in a valid normal graph for a LETS structure of 
a variable-regular Tanner graph with variable degree $d_\mathrm{v}$, the $m$ edges of the tree will have to be connected to $m$ nodes
of $\mathcal{S}$ with degree strictly less than $d_\mathrm{v}$. In addition, the degree $m$ of the root node must be at least 2 and 
at most $d_\mathrm{v}$. 

The {\em path} expansion of ${\cal S}$ is a LETS structure ${\cal S}'$ of size $a+m$, that
is constructed by appending a  path of length $m+1$ to ${\cal S}$. The first and the last nodes of the  path are common with ${\cal S}$, and can be identical, in that case, the  path is closed.
Fig. \ref{fig:expans}(b), shows the {\em path} expansion of $\mathcal{S}$ using open and closed paths of length $m+1$. 
It is clear that for an open-{\em path} expansion, the degrees of the two nodes that are common with ${\cal S}$ must be strictly less than $d_{\mathrm{v}}$ (in $G({\cal S})$), and for the closed-{\em path} expansion,
the degree of the one common node must be strictly less than $d_{\mathrm{v}}-1$ (in $G({\cal S})$). We use the notations $pa^o_m$ and $pa^c_m$ for open and closed paths of length $m+1$, respectively.  
The notation $pa_m$ is used to include both open and closed paths. 

In Fig. \ref{fig:expans}(c), the expansion of a LETS structure $\mathcal{S}$ using a lollipop walk of length $m+1$ is shown. 
The notation $lo^c_m$ is used for a lollipop walk of length $m+1$ ($m$ is number of the nodes added to $\mathcal{S}$), 
which consists of a cycle of length $c$ $(c \geq g/2)$ and a path of length $d$ $(d \geq 1)$.
Clearly, the common node between ${\cal S}$ and the {\em lollipop} expansion must have a degree strictly less than $d_{\mathrm{v}}$ in $G({\cal S})$.

The following proposition shows how the class of a LETS structure is changed as a result of the application of each of the three expansions.

\begin{pro}
\label{prois}
~\cite{hashemireg} Suppose that $\mathcal{S}$ is an $(a,b)$ LETS structure of variable-regular Tanner graphs with variable degree $d_\mathrm{v}$. Then, the expansion 
of $\mathcal{S}$ using $dot_m$ with $ 2 \leq m \leq min\{d_\mathrm{v},b\}$, $pa_m$ with $m \geq 2$, and $lo^c_m$ with $ m \geq g/2$, $g/2 \leq c \leq m$, will
result in LETS structure(s) in the  $(a+1,b+d_\mathrm{v}-2m)$, $(a+m, b-2+m(d_\mathrm{v}-2))$, and $(a+m, b-2+m(d_\mathrm{v}-2))$ classes, respectively.
(For the $dot_m$ and $pa_m$ expansions, the necessary condition is to have $b \geq 2$, while for $lo^c_m$, it is $b \geq 1$.)
\end{pro}

It was proved in~\cite{hashemireg} that any LETS structure of variable-regular Tanner graphs for any variable degree $d_{\mathrm{v}}$, and in any $(a,b)$ class, can be generated by applying a 
combination of $dot$,  $path$ and $lollipop$ expansions to simple cycles.
Also, in \cite{hashemireg}, given $a_{max}$ and $b_{max}$, a characterization algorithm was proposed to determine the expansion techniques needed to be applied to all the LETS structures within  
each LETS class to generate all the LETS structures in the interest range of $a \leq a_{max}$ and $b \leq b_{max}$, in a guaranteed fashion. 
The characterization of~\cite{hashemireg} is {\em minimal}, in the sense that, none of the expansion steps can be divided into smaller expansions such that the resulting new sub-structures are 
still LETSs. It is also proved in \cite{hashemireg} that any minimal characterization is based only on the expansions \textit{dot}, \textit{path} and \textit{lollipop}. 
Using the dpl characterization, an efficient exhaustive search algorithm for LETSs is also proposed in \cite{hashemireg} that requires only short-length simple cycles of the graph (prime structures) as the input. 
The maximum length of the input cycles for the dpl-based search algorithm is, in fact, provably minimal~\cite{hashemireg}. 

\section{Dpl-Based Characterization/Search of LETS Structures in Irregular LDPC Codes}
\label{sec:char/search}

In irregular Tanner graphs, for a given class of LETSs, the variety of non-isomorphic LETS structures would increase significantly compared to that of variable-regular Tanner graphs.
This is due to the variety of the degrees of variable nodes involved in LETS structures. 

\begin{ex}
Suppose that one is interested in characterizing the non-isomorphic $(a,b)$ LETS structures of  irregular Tanner graphs with variable degrees 3 and 4 in the interest range of $a \leq 5$ and $b \leq 4$.  
Table \ref{tab:projections} shows the quasi-normal graph representation of all the possible non-isomorphic LETS structures in this range. Comparing the information of this table with that of Tables VI and VIII of \cite{hashemireg}, for variable-regular graphs with $d_\mathrm{v}=3$ and $d_\mathrm{v}=4$, respectively, shows the considerable difference in the number of non-isomorphic LETS structures 
in this range for variable-regular versus irregular graphs. As an example, Table \ref{tab:projections} shows that there are 19 non-isomorphic LETS structures in the $(5,4)$ class. This number for the $(5,4)$ class in Table VIII of \cite{hashemireg} is only 2.
\begin{table*}[]
\centering
\caption{Non-isomorphic $(a,b)$ LETS structures of an irregular Tanner graph with variable degrees 3 and 4, in the range of $a \leq 5$ and $b \leq 4$}
\label{tab:projections}
\centering
\includegraphics [width=.8\textwidth]{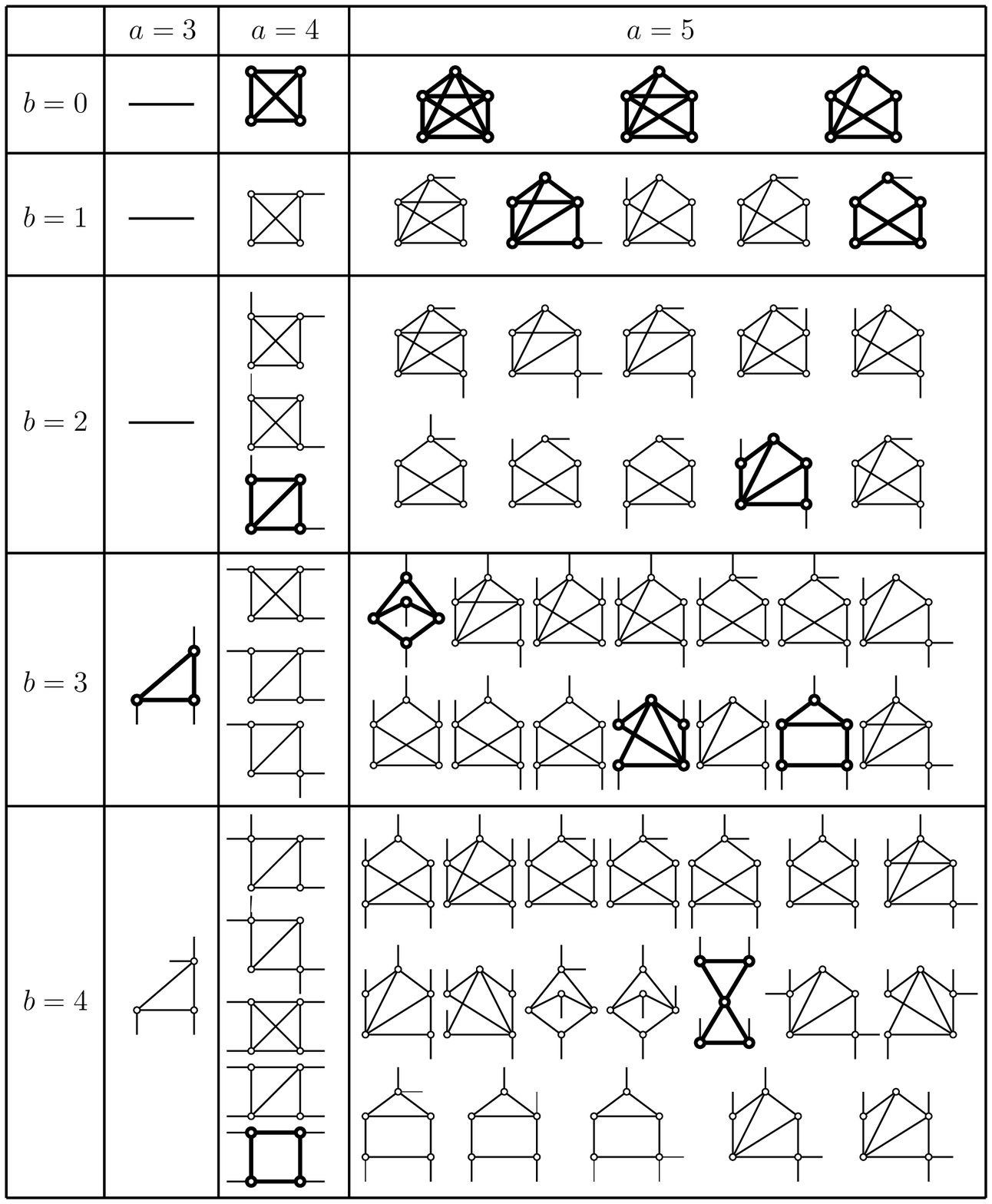}
\end{table*}
 
\end{ex}

\begin{figure}[] 
\centering
\includegraphics [width=0.45\textwidth]{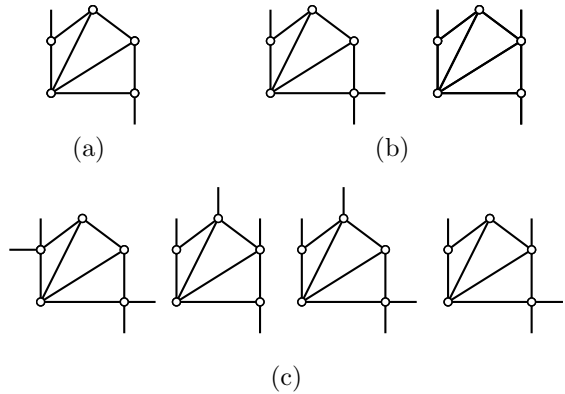}
\caption{Seven non-isomorphic LETS structures in the (a) $(5,2)$, (b) $(5,3)$ and (c) $(5,4)$ classes, all with the same normal graph .}
\label{fig:projec2}
\end{figure}

It can be seen that by increasing the range of $a$ and $b$ values or by having a larger variety of variable node degrees, the number of non-isomorphic structures increases rapidly.
An important observation, however, is that despite the large number of non-isomorphic quasi-normal graphs in each class, they are all projected to only a few normal graphs.
In Table \ref{tab:projections}, the non-isomorphic normal graphs, which are the projections of all the quasi-normal graphs in the table, are boldfaced. 
For example, the boldfaced graph in the $(5,2)$ class is the projection of 7 LETS structures, where one, two and four structures are in the $(5,2)$, $(5,3)$ and $(5,4)$ classes, respectively. 
These 7 LETS structures are presented in Fig. \ref{fig:projec2}. 

\subsection{Dpl characterization of LETS structures in irregular graphs}

 \label{sec:char}
 
In the following, we demonstrate, through a sequence of intermediate results, that the dpl characterization of $(a,b)$ LETS structures of variable-regular Tanner graphs of a properly selected variable degree $d_\mathrm{v}$,
over a properly chosen range $a \leq a'_{max}$ and $b \leq b'_{max}$ can be used to exhaustively cover all the normal graphs of all the non-isomorphic $(a,b)$ LETS structures of irregular Tanner graphs 
with a given degree distribution $\lambda(x)$ over any desired range of $a \leq a_{max}$ and $b \leq b_{max}$. (The values $a'_{max}$,  $b'_{max}$, and $d_\mathrm{v}$ are functions of $a_{max}$, $b_{max}$
and $\lambda(x)$.) As we subsequently show, this means that a dpl characterization can also be established for irregular graphs.

Suppose that $\mathcal{L}^a_{d_\mathrm{v}}$ is the set of all non-isomorphic LETS structures of size $a$ in the Tanner graph of a variable-regular graph with variable degree $d_\mathrm{v}$. 
It is easy to see that the structures in  $\mathcal{L}^a_{d_\mathrm{v}}$ can have $b$ values in the range $0 \leq b \leq a(d_\mathrm{v}-2)$. The following proposition establishes a relationship 
between the sets $\mathcal{L}^a_{d_\mathrm{v}}$, for different values of $d_\mathrm{v}$.

\begin{pro}
\label{pro:sub}
In the space of normal graphs, if $d_\mathrm{v} < a-1$,  then, $\mathcal{L}^a_{d_\mathrm{v}} \subset \mathcal{L}^a_{a-1}$, and if $d_\mathrm{v} > a-1$, then, $\mathcal{L}^a_{d_\mathrm{v}}=\mathcal{L}^a_{a-1}$.
\end{pro}

\begin{proof}
For the first part, consider an arbitrary element ${\cal S}$ of $\mathcal{L}^a_{d_\mathrm{v}}$. The LETS ${\cal S}$ has $a$ nodes, each node having a degree $d$ in the range $2 \leq d \leq d_\mathrm{v}$.
Since $\mathcal{L}^a_{a-1}$ includes {\em all} the LETS structures with $a$ nodes, where each node can have a degree in the range $[2,a-1]$, and since $d_\mathrm{v} < a-1$, the structure ${\cal S}$ is also in $\mathcal{L}^a_{a-1}$, and thus 
$\mathcal{L}^a_{d_\mathrm{v}} \subset \mathcal{L}^a_{a-1}$.

For the second part, following similar steps as in the proof of the first part, it can be shown that if $a-1 < d_\mathrm{v}$, then $\mathcal{L}^a_{a-1} \subset \mathcal{L}^a_{d_\mathrm{v}}$. 
Now, consider a structure ${\cal S}$ in $\mathcal{L}^a_{d_\mathrm{v}}$ that is not in  $\mathcal{L}^a_{a-1}$. Structure ${\cal S}$ must then have at least one node $v$ with degree strictly larger than $a-1$.
This is, however, impossible as node $v$ must be connected to at least $a$ other nodes, while there are only $a-1$ other nodes in ${\cal S}$. We thus have $\mathcal{L}^a_{d_\mathrm{v}}=\mathcal{L}^a_{a-1}$.
\end{proof}

It is important to note that even if two sets of LETS structures with different variable degrees are identical (in the space of normal graphs), they still correspond to different sets of classes. This is explained in the following examples.

\begin{ex}
Fig. \ref{fig:classes4} shows all the non-isomorphic LETS structures with size $a=4$ in a variable-regular Tanner graph with $d_\mathrm{v}=3$ and $g=6$. 
Based on the second part of Proposition \ref{pro:sub}, these structures are also all the non-isomorphic LETS structures with size $a=4$ in a variable-regular Tanner graph with $d_\mathrm{v} > 3$ and $g=6$. 
For variable-regular graphs with $d_v=3$ and $g=6$, the structures in Figs. \ref{fig:classes4} (a), (b) and (c) are the only structures in the $(4,0)$, $(4,2)$ and $(4,4)$ classes, respectively (see Table VI in \cite{hashemireg}). 
The same structures,  for variable-regular graphs with $d_v=4$ and $g=6$, are the only structures in the $(4,4)$, $(4,6)$ and $(4,8)$ classes, respectively (see Table VIII in \cite{hashemireg}).

\begin{figure}[] 
\centering
\includegraphics [width=0.3\textwidth]{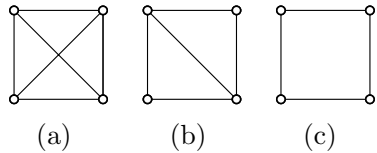}
\caption{All the non-isomorphic LETS structures with size $a=4$ in a variable-regular Tanner graph with $d_\mathrm{v} \geq 3$.}
\label{fig:classes4}
\end{figure}
\end{ex}

\begin{ex}
Fig. \ref{fig:classes5} provides all the non-isomorphic LETS structures with size $a=5$ in a variable-regular Tanner graph with $d_\mathrm{v}=4$ and $g=6$. 
Based on the second part of Proposition \ref{pro:sub}, these structures are also all the non-isomorphic LETS structures with size $a=5$ in a variable-regular 
Tanner graph with $d_\mathrm{v} > 4$ and $g=6$.  For example, $\mathcal{L}^5_5 = \mathcal{L}^5_4$. Moreover, based on the first part of Proposition \ref{pro:sub}, $\mathcal{L}^5_3 \subset \mathcal{L}^5_4$.
Table \ref{tab:classes5} shows the classes of these structures in variable-regular Tanner graphs with $d_\mathrm{v}=3, 4, 5$, and $g=6$. 
 
\begin{figure}[] 
\centering
\includegraphics [width=0.6\textwidth]{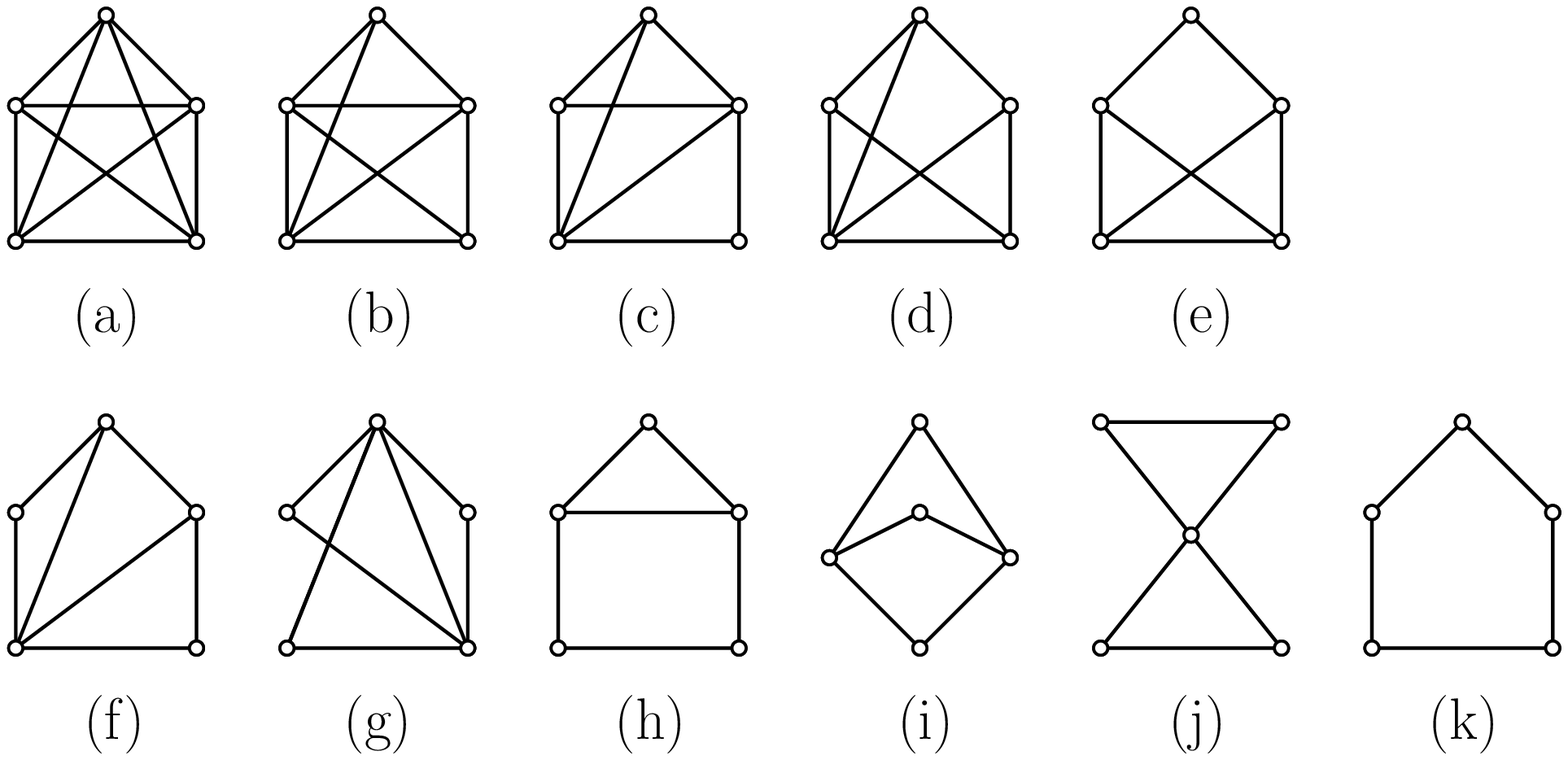}
\caption{All the non-isomorphic LETS structures with size $a=5$ in a variable-regular Tanner graph with $d_\mathrm{v} \geq 4$.}
\label{fig:classes5}
\end{figure}

\begin{table}[]
\centering
\setlength{\tabcolsep}{2pt}
\caption{Classes of non-isomorphic LETS structures in Fig. \ref{fig:classes5} for variable-regular graphs with $d_\mathrm{v}=3,4$ and $5$}
\label{tab:classes5}
\begin{tabular}{||c|c|c|c|c|c|c|c|c|c|c|c|| }
\cline{1-12}
Fig. \ref{fig:classes5}&(a)&(b)&(c)&(d)&(e)&(f)&(g)&(h)&(i)&(j)&(k)\\
\cline{1-12}
$d_\mathrm{v}=3$&-&-&-&-&(5,1)&-&-&(5,3)&(5,3)&-&(5,5)\\
\cline{1-12}
$d_\mathrm{v}=4$&(5,0)&(5,2)&(5,4)&(5,4)&(5,6)&(5,6)&(5,6)&(5,8)&(5,8)&(5,8)&(5,10)\\
\cline{1-12}
$d_\mathrm{v}=5$&(5,5)&(5,7)&(5,9)&(5,9)&(5,11)&(5,11)&(5,11)&(5,13)&(5,13)&(5,13)&(5,15)\\
\cline{1-12}
\end{tabular}
\end{table}
\end{ex}

The following proposition explains how different LETS structures of an irregular Tanner graph, corresponding to different quasi-normal graphs, are mapped (projected) to normal graphs of 
LETS structures of a variable-regular Tanner graph.
 
\begin{pro}
\label{pro:surj}
Any (quasi-normal graph of a) LETS structure $\mathcal{S}$ in the $(a,b)$ class of an irregular Tanner graph with variable degree distribution $\lambda(x)$ is mapped (via a surjective mapping) to a (normal graph of a) LETS structure in the $(a,b')$ class of variable-regular Tanner graphs with variable degree $d_\mathrm{v} = f$, where $f$ is the largest variable degree in $\lambda(x)$ strictly less than $a$ and $b'=a \times f +b - \sum\limits_{i=1}^{a} d_{v_{i}}$, with $v_{i}, i = 1, \ldots, a$, being the variable nodes of $\mathcal{S}$.
\end{pro}

\begin{proof}
The normal graph representation of $\mathcal{S}$ has $a$ nodes, where each of them is connected to at most $f$ other nodes in $\mathcal{S}$ ($f$ is the largest variable degree in $\lambda(x)$ strictly less than $a$).  
Based on Proposition~\ref{pro:sub}, this normal graph is thus a LETS in variable-regular graphs with any variable degree $d_\mathrm{v} \geq f$. Selecting the minimum variable degree in this range, i.e., $d_\mathrm{v}= f$, 
we can easily find the class of this structure using Lemma \ref{lem:eb}. For this, we note that $\mathcal{S}$ has $e= (\sum\limits_{i=1}^{a} d_{v_{i}}-b)/2$ edges. Based on Lemma \ref{lem:eb}, the number of
unsatisfied check nodes of $\mathcal{S}$ in a variable-regular graph with $d_\mathrm{v}= f$ is $b'=a \times f - 2e = a \times f +b - \sum\limits_{i=1}^{a} d_{v_{i}}$.
\end{proof}

\begin{ex}
Fig. \ref{fig:surjec} shows a LETS structure in the $(4,3)$ class of an irregular graph with variable degrees $2, 3, 4$, and the process of surjective mapping to a LETS structure in the $(4,2)$ class of a variable-regular graph with $d_\mathrm{v}=3$.
\begin{figure}[] 
\centering
\includegraphics [width=0.55\textwidth]{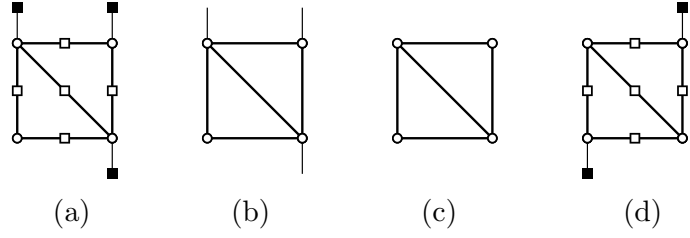}
\caption{(a) Tanner graph, and (b) quasi-normal representations of a LETS structure in the $(4,3)$ class of an irregular graph with variable degrees $2, 3, 4$; (c) normal graph, and (d) Tanner graph representations 
of a LETS structure in the $(4,2)$ class of variable-regular graphs with $d_\mathrm{v}=3$.}
\label{fig:surjec}
\end{figure}
\end{ex}

Proposition~\ref{pro:surj} describes how the LETS structures of irregular graphs are mapped to LETS structures of variable-regular graphs via the normal graph representation. 
The following theorem explains how the dpl characterization of LETS structures in variable-regular graphs can be used to characterize the LETS structures of irregular graphs (in a given range of $a$ and $b$ values, exhaustively). 
\begin{theo}
\label{theo:main}
The dpl characterization of non-isomorphic LETS structures for variable-regular graphs with $d_\mathrm{v}=t$ in $(a,b)$ classes with $a \leq a_{max}$ and $b \leq b_{max}+a_{max}(t-d_{v_{min}})$ is sufficient 
to generate the normal graphs of all the non-isomorphic $(a,b)$ LETS structures with $a \leq a_{max}$ and $b \leq b_{max}$ of irregular Tanner graphs with variable degree distribution $\lambda(x)=\sum\limits_{i=d_{v_{min}}}^{d_{v_{max}}} \lambda_i x^{i-1}$, where $t$ is the largest variable degree in $\lambda(x)$ strictly less than $a_{max}$.
\end{theo}

\begin{proof}
Proposition \ref{pro:surj} describes the mapping between the LETS structures of irregular graphs to those of variable-regular graphs. Based on this proposition, to cover the projections of all the LETS structures with $a \leq a_{max}$ of irregular graphs, the variable node degree of the variable-regular graph needs to be $d_\mathrm{v}=t$, where $t$ is the largest variable degree in $\lambda(x)$ strictly less than $a_{max}$. 
Moreover, to cover all the LETS classes of the irregular graphs in the interest range of $a \leq a_{max}$ and $b \leq b_{max}$, based on Proposition \ref{pro:surj}, the maximum value of $b'$ is $b_{max}+a_{max}(t-d_{v_{min}})$.
This is obtained by noting that, in Proposition~\ref{pro:surj}, $f=t$, and that $b'$ is maximized by setting $a$ and $b$ to their maximum values $a_{max}$ and $b_{max}$, respectively, and by minimizing $\sum\limits_{i=1}^{a_{max}} d_{v_{i}}$ through assuming that all the variable nodes in the LETS have the minimum degree $d_{v_{min}}$.
\end{proof}

\begin{ex}
\label{ex:charireg}
Based on Theorem \ref{theo:main}, to cover all the $(a,b)$ LETS structures in the range of $a \leq 7$ and $b \leq 3$, of an irregular graph with variable degrees 3, 4 and 7,  one should 
characterize all the LETS structures in a variable-regular graph with $d_\mathrm{v} = 4$ in the range of $a \leq 7$  and $b \leq 10$. 
This characterization is summarized in Table \ref{tab:4,6c3}. This representation of characterization is similar to that of \cite{hashemireg}. In Table \ref{tab:4,6c3}, columns and rows correspond to different values of $a$ and $b$, respectively. For each $(a,b)$ class of LETSs, the top entries in the table show the simple cycles that are parents of the LETS structures within that class, and the multiplicity of non-isomorphic structures in the class with those parents.    
The bottom entries show the expansion techniques applied to all the LETS structures within the class to exhaustively generate all the LETS structures in the range $a \leq 7$ and $b \leq 10$. The boldfaced entries are the prime structures required to characterize all the LETS structures in the table.
\begin{table}[]
\centering
\setlength{\tabcolsep}{.7pt}
\caption{Characterization of Non-Isomorphic LETS Structures of $(a,b)$ Classes for Variable-regular Graphs with $d_\mathrm{v}=4$ and $g=6$ for $a \leq a_{max}=7$ and $b \leq b_{max}=10$}
\label{tab:4,6c3}
\begin{tabular}{||c|c| c| c| c| c|| }
\cline{1-6}
&$a = 3$&$a = 4$&$a = 5$&$a = 6$&$a = 7$\\
\cline{1-6}
b = 0&-&-
&$  \begin{array}{@{}c@{}}s_3(1)\\ \hdashline -\end{array}  $
&$  \begin{array}{@{}c@{}}s_3(1)\\ \hdashline -\end{array}  $
& $  \begin{array}{@{}c@{}}s_3(2)\\ \hdashline -\end{array}  $
\\
\cline{1-6}
b = 1&-&-&-&-&-\\
\cline{1-6}
b = 2&-&-&
$  \begin{array}{@{}c@{}}s_3(1)\\ \hdashline dot\end{array}  $&
$  \begin{array}{@{}c@{}}s_3(3)\\ \hdashline dot\end{array}  $
& $  \begin{array}{@{}c@{}}s_3(9)\\ \hdashline -\end{array}  $
 \\
\cline{1-6}
b = 3&-&-&-&-&-\\
\cline{1-6}
b = 4
&-
&$  \begin{array}{@{}c@{}}s_3(1)\\ \hdashline dot\end{array}  $
&$  \begin{array}{@{}c@{}}s_3(2)\\ \hdashline dot\end{array}  $
&$  \begin{array}{@{}c@{}}s_3(7)\\ \hdashline dot\end{array}  $
& $  \begin{array}{@{}c@{}}s_3(27),s_4(1)\\ \hdashline -\end{array}  $
 \\
\cline{1-6}
b = 5&-&-&-&-&-\\
\cline{1-6}
b = 6
&$  \begin{array}{@{}c@{}}\boldsymbol{s_3(1)}\\ \hdashline dot,pa_2\\pa_3,lo^3_3\end{array}  $
&$  \begin{array}{@{}c@{}}s_3(1)\\ \hdashline dot,pa_2\\pa_3\end{array}  $
&$  \begin{array}{@{}c@{}}s_3(3)\\ \hdashline dot\end{array}  $
&$  \begin{array}{@{}c@{}}s_3(10),s_4(1)\\ \hdashline dot \end{array}  $
& $  \begin{array}{@{}c@{}}s_3(43),s_4(1) \\ \hdashline -\end{array}  $
\\
\cline{1-6}
b = 7&-&-&-&-&-\\
\cline{1-6}
b = 8
&-
&$  \begin{array}{@{}c@{}}\boldsymbol{s_4(1)}\\ \hdashline dot,pa_2\end{array}  $
&$  \begin{array}{@{}c@{}}s_3(2),s_4(1)\\ \hdashline dot,pa_2\end{array}  $
&$  \begin{array}{@{}c@{}}s_3(8),s_4(2)\\ \hdashline dot \end{array}  $
& $  \begin{array}{@{}c@{}}s_3(41),s_4(3)\\ \hdashline -\end{array}  $
 \\
\cline{1-6}
b = 9&-&-&-&-&-\\
\cline{1-6}
b = 10
&-
&-
&$  \begin{array}{@{}c@{}}\\ \hdashline -\end{array}  $
&$  \begin{array}{@{}c@{}}s_3(3),s_4(2)\\ \hdashline dot \end{array}  $
& $  \begin{array}{@{}c@{}}s_3(21),s_4(6)\\ \hdashline - \end{array}  $
 \\
\cline{1-6}
\end{tabular}
\end{table}
Note that to cover this rather small range of $a$ and $b$ values for the irregular graph, one needs to cover a wide range of $b$ values for the corresponding variable-regular graph.
\end{ex}

\subsection{Dpl-based search algorithm for irregular graphs}
\label{sec:search1}

In~\cite{hashemireg}, for variable-regular graphs, the dpl characterization of LETS structures was used as a road-map for the dpl-based search algorithm to find all the instances of 
LETS structures in any interest range of $a$ and $b$ values in a guaranteed fashion. The search algorithm of \cite{hashemireg} starts by the enumeration of simple cycles. 
These are the cycles that are identified in the characterization table as the required prime structures. After the enumeration of all the instances of a simple cycle, 
these instances are expanded recursively to find instances of other LETS structures up to size $a_{max}$. In each step, after finding a new LETS, the indices of its variable nodes should be saved for subsequent expansions in the next step.
The expansion techniques identified by the characterization table should be applied to all the instances of LETS structures in the corresponding classes. 

In Subsection \ref{sec:char}, it was shown that the characterization table for variable-regular graphs with variable degree $d_\mathrm{v}=t$ in the range of $a \leq a_{max}$ and $b \leq b_{max}+a_{max}(t-d_{v_{min}})$ can be used to characterize the $(a,b)$ LETS structures of irregular graphs with degree distribution $\lambda(x)=\sum\limits_{i=d_{v_{min}}}^{d_{v_{max}}} \lambda_i x^{i-1}$ in the range of $a \leq a_{max}$ and $b \leq b_{max}$, where $t$ is the largest variable degree less than $a_{max}$ in $\lambda(x)$. 
Unlike the case for variable-regular graphs, in irregular Tanner graphs, however, there is no one-to-one correspondence between a LETS structure and its normal graph, i.e., a normal graph can be 
the projection of multiple LETS structures in different classes. In the search algorithm for irregular graphs, therefore, for each LETS, in addition to the index of its variable nodes, we need to also keep track of the class of its normal 
graph in variable-regular graphs with $d_\mathrm{v}=t$. Otherwise, the search follows the exact similar steps as in the dpl-based search of \cite{hashemireg}. More details are provided in the following.

The dpl-based search algorithm of irregular graphs starts by the enumeration of simple cycles (as identified in the characterization table for the corresponding variable-regular graph with  $d_\mathrm{v}=t$). 
For different instances of a simple cycle of a given length, based on the degrees of the variable nodes in the cycles, the cycles would belong to different classes, all with the same $a$ but different $b$ values. 
In addition to saving the information of each instance (indices of its variable nodes), the class of its normal graph in the characterization table for variable-regular graphs with  $d_\mathrm{v}=t$ should be saved as well. 
After the enumeration of all instances of a simple cycle, these instances are expanded recursively to find instances of other LETS structures up to size $a_{max}$ following the dpl-based search algorithm (Algorithm 4 of \cite{hashemireg}).
It is important to note that, in the search process, no matter what the actual value of $b$ of an instance of a LETS is, as long as its normal graph is in the $(a,b)$ class of variable-regular graphs with $d_\mathrm{v}=t$, 
the characterization table for the variable-regular graph shows which expansions should be applied to that instance. 
Also, no matter what the class of the resultant instance is, one only needs to keep track of the class of its normal graph (in the characterization table for variable-regular graphs with $d_\mathrm{v}=t$).
This is done based on Proposition~\ref{prois}.  

\begin{ex}
Suppose that in an irregular graph with variable degrees $3,4$ and $7$, there are a set of instances of LETSs in different classes with $a=4$ and $b=2,3,4,6,7$ and $8$. 
Also suppose that all these instances are projected to the normal graph in the $(4,4)$ class of Table \ref{tab:4,6c3}. Applying $dot_2$ to all these instances may result in new instances with $a=5$ and, for example, $b=1,2,5,7$, and $8$. 
However, based on Proposition~\ref{prois}, the normal graph of all the new instances will be in the $(5,4)$ class of Table \ref{tab:4,6c3}. Therefore, no matter what the classes of new instances are, 
the class $(5,4)$ should be saved for each of them for the next step. This class is the one that is used to determine the expansions that will have to be applied to these instances in the next step (based on the lower entries 
of the $(5,4)$ cell in Table \ref{tab:4,6c3}).
\end{ex}

\begin{ex}
\label{exsw}
Consider the case of Example \ref{ex:charireg}. The LETS structures of irregular graphs in the $(5,4)$ class are projected to normal graphs in classes $(5,4)$, $(5,6)$ and $(5,8)$ of Table \ref{tab:4,6c3}.
One should thus apply  \{\textit{$dot$}\}, \{\textit{$dot$}\} and \{\textit{$dot,pa_2$}\} expansions to the instances of corresponding LETS structures, respectively.
\end{ex}

\section{Efficient exhaustive Search Algorithm for LETSs of Irregular LDPC Codes}
\label{sec:search}

The search algorithm for LETSs of irregular graphs described in Section \ref{sec:char/search} can have two major problems, both occurring when the size of the 
characterization table for the variable-regular graphs with $d_\mathrm{v}=t$, where $t$ is the largest variable degree strictly less than $a_{max}$, happens to become too large.
This can happen, in the cases that the value of $t$ is relatively large (for example $t \geq 8$), or in the cases where the interest range of $a$  is relatively large (for example $a_{max} \geq 10$),
or in the cases where variable nodes with small degrees (for example, degree-2 variable nodes) are present. Under such circumstances, the range of $a$ and $b$ values covered in the characterization table of 
variable-regular graphs with $d_\mathrm{v}=t$ plus the number of non-isomorphic LETS structures in the table will quickly increase.  For example, there are more than 
nine million non-isomorphic LETS structures in the classes with $a=10$ for variable-regular graphs with $d_\mathrm{v}=8$ ($|\mathcal{L}^{10}_{8}|=9,545,887$).
The first problem resulting from the large size of the characterization table is the excessive time that it takes the characterization algorithm of~\cite{hashemireg} to generate the table 
in addition to the large amount of memory that is required to store the information of the characterization table. The second problem is the inefficiency of the search algorithm corresponding to a large
characterization table, in the sense that, one may have to enumerate LETS structures with large values of $b$ (and large multiplicities) that are not of direct interest but are parents of LETS structures of interest.
  
To overcome the above problems in circumstances just explained, we use a different approach to characterize/search LETSs of irregular codes. Rather than relying on the normal graph representation through the characterization table of a variable-regular graph, in the new approach, we focus on the class of LETS structures, i.e., rather than tracking the class of the normal graph within the characterization table of the corresponding variable-regular graph, we 
concern ourselves with the class of the LETS itself. For each class of LETS structures, we then identify {\em all} the possible expansions from the set of {\em dot}, {\em path} and {\em lollipop} expansions 
that can be applied to LETS structures in that class and eventually (after successive application of one or more expansions) result in an $(a,b)$ LETS structure with $a \leq a_{max}$ and $b \leq b_{max}$.
As an example, consider the scenario of Example \ref{ex:charireg}. Based on the previous approach, for any LETS structure of the irregular graphs which is projected to one of the two normal graphs in
the $(5,4)$ class of Table~\ref{tab:4,6c3}, regardless of the actual class of the LETS structure, the {\em dot} expansion is applied. In the new approach, however, the expansions are decided by the class of the LETS structure
itself. For example, for a LETS structure in the $(5,4)$ class, all the possible expansions that can eventually result in a LETS in the range of $a \leq 7$ and $b \leq 3$ are applied. These are {\em dot} and ${pa}_2$ expansions.
(Note that based on Example~\ref{exsw}, the LETS structures in the class of $(5,4)$ can be projected into normal graphs in three different classes of Table   \ref{tab:4,6c3}, including the $(5,4)$ class.)
The application of this approach recursively and starting from all the simple cycles with size from $g/2$ up to $a_{max}$ can provide us with an exhaustive list of all LETS structures of an irregular code within any desired range of $a \leq a_{max}$ and $b \leq b_{max}$. The indiscriminate application of this approach, however, is inefficient in both characterization and search of LETS structures. The reason is that many of the structures generated initially or in the intermediate stages of this process may not eventually reach any LETS structure within the range of interest. 
To overcome this problem while maintaining the exhaustiveness of the characterization/search, we identify the expansions for each class through a backward recursion. We start from LETS structures of size $a_{max}$ with $b \leq b_{max}$, and then find out how these structures can be possibly constructed 
in a recursive fashion starting from simple cycles using the three expansions. The first step is to find out all the possible candidates that can lead to the target LETS structures through a single expansion step and then 
backtrack this recursively until one reaches simple cycles. 
  
In the characterization/search of LETSs in variable-regular graphs, to find (cover) all the LETSs in the range of $a \leq a_{max}$ and $b \leq b_{max}$, one should sometimes include $(a,b)$ LETS structures 
with their $b$ values outside the range of interest, i.e., $b_{max} < b \leq b'_{max}$.  In \cite{hashemireg}, it was proved that the value of $b'_{max}$ selected by the dpl characterization/search is minimal. 
In this work, for irregular graphs, for any given value of $a$ in the range $ g/2 \leq a \leq a_{max}$, we derive an upper bound $b^a_{max}$, on the $b$ values for the classes of $(a,b)$ LETS structures. 
This upper bound determines all the $(a,b)$ classes of the same size $a$, i.e., $(a,b)$ classes with $b \leq b^a_{max}$, that need to be covered in the characterization/search such that by the application of all the 
possible expansions to the included classes, starting from simple cycles, one can obtain all the LETS structures of the irregular graph within the range of interest $a \leq a_{max}$ and $b \leq b_{max}$, exhaustively. 
The upper bound is derived recursively by finding $b^a_{max}$ as a function of $b^{a+1}_{max}$ with the initial condition that $b^{a_{max}}_{max}$ is equal to $b_{max}$.

The following lemma imposes a limit on the maximum degree of a variable node that can be involved in a LETS structure given the interest range
of $a \leq a_{max}$ and $b \leq b_{max}$. This will help to further constrain the space of search, and reduce its complexity.

\begin{lem}
\label{lem3}
Any variable node $v$ involved in an $(a,b)$ LETS structure of an irregular graph in the range of $a \leq a_{max}$ and $b \leq b_{max}$, has a degree $deg(v) < a_{max}+b_{max}$. 
\end{lem}

\begin{proof}
Consider a variable node $v$ with degree $deg(v) \geq  a_{max}+b_{max}$ in an $(a,b)$ LETS structure with $a \leq a_{max}$. This means that $v$ is connected to at most $a_{max} - 1$ variable nodes 
within the LETS structure. Since $deg(v) \geq  a_{max}+b_{max}$, this implies that $v$ has at least $b_{max}+1$ unsatisfied check nodes in its neighbourhood, which is in contradiction 
with the LETS structure having a $b$ value of at most $b_{max}$.
\end{proof}

The next lemma can, in some cases, further limit the search space.
 

\begin{lem}
\label{pro:cycle}
In an irregular Tanner graph with variable node degree distribution $\lambda(x)$, among the LETS structures with the same size $a$, the simple cycle, consisting of variable nodes all with degree $d_{v_{max}}$, has the largest $b$ value, equal to $a(d_{v_{max}} - 2)$.
Moreover, for such an irregular graph, among $(a,b)$ LETS structures with $a \leq a_{max}$ and $b \leq b_{max}$, the simple cycle of size $a$ in the $(a, a(\eta-2))$ 
class has the largest $b$ value, where $\eta$ is the largest variable degree in $\lambda(x)$ strictly smaller than $a_{max}+b_{max}$.
\end{lem}

As an intermediate step to deriving the upper bounds, in the following proposition, we first determine how the class of a LETS structure of an irregular graph changes as it is expanded by one of the three expansion techniques.
In this proposition, we use the notation $dot_m^k$ to denote a dot expansion with $m$ edges and a root node with degree $k$.

\begin{pro}
\label{pro:dot,pa,lo}
Consider an $(a,b)$ LETS structure ${\cal S}$ in an irregular Tanner graph.  
The application of $dot^{k}_m$ ($2 \leq m \leq k$), $pa_m$ ($m \geq 2$), and $lo^c_m$ ($m \geq g/2, g/2 \leq c \leq m$) to ${\cal S}$ 
will result in LETS structure(s) in the  $(a+1,b+k-2m)$, $(a+m,b-2+\sum\limits_{i=1}^{m} (d_{v_{i}}-2))$, and $(a+m,b-2+\sum\limits_{i=1}^{m} (d_{v_{i}}-2))$ classes, respectively,
where $d_{v_{i}}, i = 1, \ldots, m$, are the degrees of the variable nodes in the expansion.

\end{pro}

\begin{proof}
Similar to the proof of Proposition~\ref{prois}, as given in \cite{hashemireg}.
\end{proof}

\begin{pro}
\label{corgk}
In an irregular Tanner graph with variable node degree distribution $\lambda(x)$, the application
of $dot^{k}_m$ ($2 \leq m \leq k$), $pa_m$ ($m \geq 2$), and $lo^c_m$ ($m \geq g/2, g/2 \leq c \leq m$) to an $(a,b)$ LETS structure results in an $(a',b')$ LETS structure with
minimum possible value of $b'$ equal to  $(b+\min\{z-2a,-y\})^+$, $(b-2+m(d_{v_{min}}-2))^+$, and $(b-2+(m-1)(d_{v_{min}}-2) + d'-2)^+$, respectively, where $(f)^+=\max\{f,0\}$,
$y$ is the largest variable degree in $\lambda(x)$ less than or equal to $a$, $z$ is the smallest variable degree in $\lambda(x)$ strictly larger than $a$,
and $d'$ is equal to $d_{v_{min}}$, if $d_{v_{min}} > 2$, and is equal to the smallest variable degree strictly larger than $2$, if $d_{v_{min}} = 2$.
In particular, if $d_{v_{min}} = 2$, the minimum values of $b'$ for $pa_m$ and $lo^c_m$ are $(b-2)^+$ and $(b+d'-4)^+$, respectively, where
$d'$ is equal to the smallest variable degree strictly larger than $2$.
\end{pro}

\begin{proof}
For the $dot^{k}_m$ expansion, using Proposition~\ref{pro:dot,pa,lo}, we have $b'=b+k -2m$, and we note that $m \leq \min\{a,k\}$. We consider two cases: (i) $a < k$ and (ii) $a \geq k$. 
For Case (i), we have $m \leq a$, and the minimum value of $b'$ is obtained when $m$ has its maximum value in this interval, i.e., $m=a$, and $k$ has its smallest value in the interval $k > a$, i.e., $k=z$.
This results in $b' = b+z -2a$. For Case (ii), we have $m \leq k$, and the minimum value of $b'$ is obtained when $m$ has its maximum value in this interval, i.e., $m=k$. This results in $b'=b-k$, which in turn is minimized if 
$k$ takes its largest value in the interval $k \leq a$, which is $y$. This results in $b'=b-y$. Combining the results of Cases (i) and (ii), we obtain $b'= (b+\min\{z-2a,-y\})^+$, where $(\cdot)^+$ simply 
indicates that the $b$ value cannot be negative. For the $pa_m$ expansion, the minimum is attained when all the $m$ nodes in the expansion have degree $d_{v_{min}}$. For the $lo^c_m$, the minimum is resulted
when all the nodes in the expansion have degree $d_{v_{min}}$, with the exception being when $d_{v_{min}} = 2$, in which case, one node needs
to have a degree equal to the smallest variable degree strictly larger than $2$.
\end{proof}

Based on Proposition~\ref{corgk}, it can be seen that $pa_m$ and $lo_m^c$ expansions can cause the decrease of at most $2$ and $1$ in the $b$ value of a LETS structure, respectively.
The $dot$ expansion, however, can cause larger decreases in the $b$ value of a LETS structure.

The following theorem establishes the upper bounds $b^a_{max}, g/2 \leq a \leq a_{max}$, through a recursive relationship.
 
\begin{theo}
\label{theo:sear}
Suppose that we are interested in generating all the $(a,b)$ LETS structures of an irregular Tanner graph with variable node degree distribution $\lambda(x)$ and girth $g$ within the range $a \leq a_{max}$ and $b \leq b_{max}$.
For this, consider an approach that starts from simple cycles $s_k, k=g/2, \ldots, a_{max}$, and recursively applies all the possible $dot$, $path$ and $lollipop$ expansions to any generated LETS structure. 
Such an approach will exhaustively generate all the LETS structures in the range of interest, if for any size $a$ in the range $g/2 \leq a \leq a_{max}$, the approach is constrained to only find
$(a,b)$ LETS structures with $b$ values satisfying $b \leq b^a_{max}$, where $ b^a_{max}$ values are obtained through the recursion 
\begin{equation}
\label{eqkd}
b^a_{max}=\min\{b^{a+1}_{max}+ max \{y, 2a-z\}\:,\:a (\eta-2)\}\:.
\end{equation}
In~(\ref{eqkd}), $y$ is the largest variable degree in $\lambda(x)$ less than or equal to $a$, $z$ is the smallest variable degree in $\lambda(x)$ strictly larger than $a$ and strictly smaller than $a_{max}+b_{max}$, 
and $\eta$ is the largest variable degree in $\lambda(x)$ strictly smaller than $a_{max}+b_{max}$. 
The initial condition for recursion is $b^{a_{max}}_{max} = b_{max}$.
\end{theo}

\begin{proof}
Based on Proposition~\ref{corgk}, it is clear that the largest decrease in the value of $b$ by increasing $a$ through the three expansions is caused by the $dot$ expansion. In fact,
it is easy to see that the recursion $b^a_{max} = b^{a+1}_{max} +1$, along with the initial condition $b^{a_{max}}_{max} = b_{max}$, cover the range of $b$ values required for exhaustive search
based on $path$ and $lollipop$ expansions. Focusing on $dot$ then, based on Proposition~\ref{corgk}, it is clear that the largest decrease in the $b$ value by the application of $dot$ is $- \min\{z-2a,-y\}$, 
or equivalently, $\max\{2a-z,y\}$. The proof is then completed by combining this with Lemmas~\ref{lem3} and \ref{pro:cycle}.
\end{proof}

We note that Theorem~\ref{theo:sear} provides a new dpl-based exhaustive characterization/search for LETS structures of irregular codes (compared to what was presented in Section~\ref{sec:char/search}).
We also note that the upper bounds derived based on Theorem~\ref{theo:sear} may not be necessarily tight, i.e., one may be able to find smaller bounds that still result in an exhaustive coverage of the 
LETS structures of interest, thus, further reducing the complexity of the search. One can also perform the search in a smaller space, by reducing the upper bounds, but
possibly at the expense of sacrificing the exhaustiveness of the search. In fact, based on our experimental results, we propose the following upper bounds as a lower complexity alternative to (\ref{eqkd}):
\begin{equation}
\label{eqif}
b^{a}_{max}=\min\{b^{a+1}_{max}+  max \{y, 2a-z\}-2\:,\:a(\eta-2)\}\:.
\end{equation}

Although in our extensive simulations, presented in Section  \ref{sec:numerical}, the upper bounds (\ref{eqif}) have always resulted in an exhaustive search, we have not been able to prove this, in general.

Given the upper bounds $b^{g/2}_{max}, \dots, b^{a_{max}}_{max}$, Algorithm~\ref{alg1} provides a pseudo code for finding the list of expansion techniques that are
required to be applied to all the non-isomorphic structures in each $(a,b)$ LETS class. These expansions are stored in the $(a,b)$ entry of table $\mathcal{EX}$, $\mathcal{EX}_{(a,b)}$.
Based on Algorithm~\ref{alg1}, the expansion $dot$ can be applied to all the $(a,b)$ classes with $a \leq a_{max}-1$. Also, $pa_m$ and $lo_m^c$ can be applied to all the $(a,b)$ classes with $a \leq a_{max}-m$.
The only constraint for using an expansion technique is that the $b$ value(s) of the new LETS structure(s) need to remain in the range identified by the upper bounds  $b^{g/2}_{max}, \dots, b^{a_{max}}_{max}$.
The results of Proposition~\ref{pro:dot,pa,lo} are used to impose this constraint. 

\begin{algorithm}
 \caption{Finding the expansion techniques $\mathcal{EX}_{(a,b)}$ for the $(a,b)$ classes of LETS structures, $g/2 \leq a \leq a_{max}$,  $1 \leq b \leq b_{max}$, for an irregular Tanner graph with girth $g$.}
\label{alg1}
 \begin{algorithmic} [1] 
\State \textbf{Inputs:} $a_{max}, b^{g/2}_{max}, \dots, b^{a_{max}}_{max}=b_{max}, g$.
\State \textbf{Initializations:} $a=a_{max}-1$.
\While {$a \geq g/2$}
\State  $\mathcal{EX}_{(a,b)} \gets dot,~\forall~b \in \{2,\dots, b^a_{max}\}$.
\For {$b = 1, \ldots, b^a_{max}$}
\State $m=2$.
\While{$a+m \leq a_{max}$}
\If {$0 \leq b-2 \leq b^{a+m}_{max}$}
\State $\mathcal{EX}_{(a,b)}\gets pa_m$.
\EndIf
\If {$m \geq g/2$ and $ b-1 \leq b^{a+m}_{max}$}
\For{$c=g/2,\dots, m$}
\State $\mathcal{EX}_{(a,b)}\gets lo_m^c$.
\EndFor
\EndIf
\State $m=m+1$.
\EndWhile
\EndFor
\State $a=a-1$.
\EndWhile
\State \textbf{Output:} $\mathcal{EX}$.
\end{algorithmic}
 \end{algorithm}

\begin{ex}
The outcome of Algorithm \ref{alg1} is presented in Table  \ref{tab:4,6c4} for the case of $a_{max} \leq 7$, $b_{max} \leq 2$, and an irregular graph with variable node degrees $\{2,3,5, 10\}$ and $g=6$. 
\begin{table}[]
\centering
\setlength{\tabcolsep}{.5pt}
\caption{Expansions Required for $(a,b)$ Classes of Irregular Graphs with Variable Degrees $2,3,5,10$, and $g=6$ for $a \leq a_{max}=7$ and $b \leq b_{max}=2$}
\label{tab:4,6c4}
\begin{tabular}{||c|c| c| c| c| c|| }
\cline{1-6}
&$a = 3$&$a = 4$&$a = 5$&$a = 6$&$a = 7$\\
\cline{1-6}
b = 0&-&-&-&-&-\\
\cline{1-6}
b = 1
&$  \begin{array}{@{}c@{}} lo_3,lo_4\end{array}$
&$  \begin{array}{@{}c@{}} lo_3\end{array}$
&-&-&-\\
\cline{1-6}
b = 2
&$  \begin{array}{@{}c@{}} dot,pa_2,pa_3,pa_4,lo_3,lo_4\end{array} $
&$  \begin{array}{@{}c@{}} dot,pa_2,pa_3,lo_3\end{array} $
&$  \begin{array}{@{}c@{}} dot,pa_2\end{array} $
&$  \begin{array}{@{}c@{}} dot\end{array}  $
&-
\\
\cline{1-6}
\hline
\hline
b = 3
&$  \begin{array}{@{}c@{}} dot,pa_2,pa_3,pa_4,lo_3,lo_4\end{array} $
&$  \begin{array}{@{}c@{}} dot,pa_2,pa_3,lo_3\end{array} $
&$  \begin{array}{@{}c@{}} dot,pa_2\end{array} $
&$  \begin{array}{@{}c@{}} dot\end{array}  $
\\
\cline{1-5}
b = 4
&$  \begin{array}{@{}c@{}} dot,pa_2,pa_3,pa_4,lo_3\end{array} $
&$  \begin{array}{@{}c@{}} dot,pa_2,pa_3\end{array} $
&$  \begin{array}{@{}c@{}} dot,pa_2\end{array} $
&$  \begin{array}{@{}c@{}} dot\end{array}  $
\\
\cline{1-5}
b = 5
&$  \begin{array}{@{}c@{}} dot,pa_2,pa_3,lo_3\end{array} $
&$  \begin{array}{@{}c@{}} dot,pa_2\end{array} $
&$  \begin{array}{@{}c@{}} dot\end{array} $
&$  \begin{array}{@{}c@{}} dot\end{array}  $
\\
\cline{1-5}
b = 6
&$  \begin{array}{@{}c@{}} dot,pa_2,pa_3,lo_3\end{array} $
&$  \begin{array}{@{}c@{}} dot,pa_2\end{array} $
&$  \begin{array}{@{}c@{}} dot\end{array} $
&$  \begin{array}{@{}c@{}} dot\end{array}  $\\
\cline{1-5}
b = 7&$  \begin{array}{@{}c@{}} dot,pa_2,pa_3,lo_3\end{array} $
&$  \begin{array}{@{}c@{}} dot,pa_2\end{array} $
&$  \begin{array}{@{}c@{}} dot\end{array} $
&$  \begin{array}{@{}c@{}} dot\end{array} $\\
\cline{1-5}
b = 8&$  \begin{array}{@{}c@{}} dot,pa_2,pa_3,lo_3\end{array} $
&$  \begin{array}{@{}c@{}} dot,pa_2\end{array} $
&$  \begin{array}{@{}c@{}} dot\end{array} $
\\
\cline{1-4}
b = 9&$  \begin{array}{@{}c@{}} dot,pa_2,pa_3\end{array} $
&$  \begin{array}{@{}c@{}} dot,pa_2\end{array} $
&$  \begin{array}{@{}c@{}} dot\end{array} $\\
\cline{1-4}
b = 10&
&$  \begin{array}{@{}c@{}} dot\end{array} $&$  \begin{array}{@{}c@{}} dot\end{array} $\\
\cline{1-1}\cline{3-4}
b = 11&
&$  \begin{array}{@{}c@{}} dot\end{array} $&$  \begin{array}{@{}c@{}} dot\end{array} $\\
\cline{1-1}\cline{3-4}
b = 12&
&$  \begin{array}{@{}c@{}} dot\end{array} $&$  \begin{array}{@{}c@{}} dot\end{array} $\\
\cline{1-1}\cline{3-4}
\end{tabular}
\end{table}
These variable node degrees are used in \cite{saeedi} to design near-optimal irregular LDPC codes over binary-input additive white Gaussian noise (BIAWGN) channels. 
In Table \ref{tab:4,6c4}, the notation $dot$ is used  to represent any $dot_m^k$ expansion that results in $(a,b)$ LETS structures with $b \leq b^a_{max}$. 
Also, the notation $lo_m$ is used to represent all the $lo_m^c$ expansion techniques with different values of $c$. 
Based on Theorem \ref{theo:sear}, we have $b^7_{max}=2,~b^6_{max}=7,~b^5_{max}=12,~b^4_{max}=12,~b^3_{max}=9$.  
\end{ex}

The pseudo code of the proposed search algorithm is given in Algorithm \ref{alg2}. 
\begin{algorithm}
\small{
 \caption{{\bf (LETS Exhaustive Search)} Finds all the instances of $(a,b)$ LETS structures of an irregular Tanner graph $G$ with girth $g$, for $a \leq a_{max}$ and $b \leq b_{max}$. The inputs are 
the upper bounds $b^{g/2}_{max}, \dots, b^{a_{max}}_{max}$,  and expansion techniques $\mathcal{EX}$ provided by Algorithm \ref{alg1}. 
The output is the set ${\cal I}$, which contains all the instances of LETS structures in the interest range.}
\label{alg2}
 \begin{algorithmic} [1] 
\State \textbf{Inputs:} $G, a_{max}, \{b^{g/2}_{max}, \dots, b^{a_{max}}_{max}\}, \mathcal{EX}, g$.
\State  \textbf{Initializations:} $\mathcal{I} \gets \emptyset$.
\For {$k = g/2, \dots, a_{max}$}
\parState {$\mathcal{I}_k=\{\mathcal{I}_k^{k,0}, \dots, \mathcal{I}_k^{k,b^k_{max}}\}$= \textbf{CycSrch}$(G,k)$.}
\State $\mathcal{I}=\mathcal{I} \cup \mathcal{I}_k$.
\EndFor
\For {$k = g/2, \dots, a_{max}$}
\State {$a=k$.}
\While {$a < a_{max}$} \label{shorwhile}
\For{$b=1, \dots, b^a_{max}$}
\If {$dot \in \mathcal{EX}_{(a,b)}$}
\State $\{\mathcal{I}_{tem}^{a+1,0},\dots, \mathcal{I}_{tem}^{a+1,b^{a+1}_{max}}\}$= \textbf{DotSrch}$(G,\mathcal{I}_k^{a,b},\mathcal{I})$.
\EndIf
\For {any $pa_m \in \mathcal{EX}_{(a,b)}$}
\parState {$\{\mathcal{I}_{tem}^{a+m,0},\dots,\mathcal{I}_{tem}^{a+m,b^{a+m}_{max}}\}$= \textbf{PathSrch}$(G,\mathcal{I}_k^{a,b}, \mathcal{I}, m)$.}
\EndFor
\For {any $lo^c_m \in \mathcal{EX}_{(a,b)}$}
\parState {$\{\mathcal{II}_{tem}^{a+m,0},\dots, \mathcal{II}_{tem}^{a+m,b^{a+m}_{max}}\}$= \textbf{LolliSrch}$(G,\mathcal{I}_k^{a,b}, \mathcal{I},\mathcal{I}_c, m)$.}
\For {$s=1, \dots, b^a_{max}$}
\State $\mathcal{I}_{tem}^{a+m,s}=\mathcal{I}_{tem}^{a+m,s}\cup \mathcal{II}_{tem}^{a+m,s}$.
\EndFor
\EndFor
\For{$ t=a+1,\dots, a_{max}$}
\For{$s=1, \dots, b^t_{max}$}
\State $\mathcal{I}_k^{t,s}=\mathcal{I}_k^{t,s} \cup \mathcal{I}_{tem}^{t,s}$.
\State $\mathcal{I}=\mathcal{I} \cup \mathcal{I}_{tem}^{t,s}$.
\EndFor
\EndFor
\EndFor
\State $a=a+1$.
\EndWhile \label{khatwhile}
\EndFor
\State \textbf{Output:} $\mathcal{I}$.
\end{algorithmic}
}
 \end{algorithm}
Having the upper bounds $b^{g/2}_{max}, \dots, b^{a_{max}}_{max}$, and the expansion table $\mathcal{EX}$, as the input, the search algorithm 
starts by the enumeration of simple cycles of length up to $a_{max}$ through Routine \ref{tab:search cycle}.
\begin{routine}
\centering
\caption{{\bf (CycSrch)} Finds all the instances of simple cycles of length $k$ with $b \leq b^k_{max}$, in a given Tanner graph $G$. $\{\mathcal{I}_k^{k,0}, \dots, \mathcal{I}_k^{k,b^k_{max}}\}$= CycSrch$(G,k)$}
\label{tab:search cycle}
\begin{algorithmic}[1]
\State  \textbf{Initializations:} $\mathcal{I}_k^{k,b} \gets \emptyset,~\forall~b \leq b^k_{max}$.
\For {each variable node $v_l$ in $G$}
\For{each check node $c_i$ in the neighbourhood of $v_l$}
\parState {Find all the paths $\mathcal{PA}_{i,l}$ of length $k-1$ in $G$, starting from $c_i$ that do not contain $v_l$.} 
\EndFor
\parFor{any pair of check nodes $c_i$ and $c_j$ in the neighbourhood of $v_l$, where $i \neq j$}
\parFor {any path $pa \in \mathcal{PA}_{i,l}$, and any path $pa' \in \mathcal{PA}_{j,l}$}\label{cyc-comp}
\parIf {the two paths end with the same node and that node is their only common node}
\parState {Let $v$ and $v'$ denote the last variable nodes of $pa$ and $pa'$, respectively.}
\parIf{variable nodes in $pa \setminus v$ and $pa' \setminus v'$ do not have any common check node}
\parState {$\mathcal{S}=v_l \cup$\{set of variable nodes in $pa \cup pa'$\}.}
\If{$|\Gamma_{o}{(\mathcal{S})}| \leq b^k_{max}$}
\State $\mathcal{I}_k^{k,|\Gamma_{o}{(\mathcal{S})}|}=\mathcal{I}_k^{k,|\Gamma_{o}{(\mathcal{S})}|} \cup \{\mathcal{S}\}$.
\EndIf
\EndparIf
\EndparIf
\EndparFor
\EndparFor
\EndFor
\State \textbf{Outputs:} $\{\mathcal{I}_k^{k,0}, \dots, \mathcal{I}_k^{k,b^k_{max}}\}$.
\end{algorithmic}
\end{routine}
Note that for any $s_k$, where $g/2 \leq  k \leq a_{max}$, the number of unsatisfied check nodes of the instances should be less than or equal $b^k_{max}$, i.e., 
only simple cycles that satisfy this condition are stored for further processing. This is performed in Lines~12-14 of Routine~\ref{tab:search cycle}. After the enumeration of all instances of a simple cycle, these instances are expanded in the while loop (Lines \ref{shorwhile}-\ref{khatwhile}) to find instances of other LETS structures in the interest range. The selected expansions for each class are those stored in $\mathcal{EX}$. 
In Algorithm~\ref{alg2}, the notation $\mathcal{I}_k^{a,b}$ is used for the set of LETS instances in the $(a,b)$ class found by starting from the instances of the simple cycle of length $k$, 
and $\mathcal{I}$ is the set of all instances which are found so far in the algorithm. 
Finding the instances of LETS structures using \textit{dot}, \textit{path} and \textit{lollipop} expansion
techniques are explained in Routines \ref{tab:search $dot$}, \ref{tab:search $path$} and \ref{tab:search lolli}, respectively.
\begin{routine}
\centering
\caption{{\bf (DotSrch)} Expanding all the instances of LETS structures in the $(a,b)$ class $\mathcal{I}_k^{a,b}$ of Tanner graph $G$ using {\em dot} expansions to find a set of instances  of LETS structures of size $a+1$,
excluding the already found structures $\mathcal{I}$, and storing the rest in $\{\mathcal{I}_{tem}^{a+1,0},\dots, \mathcal{I}_{tem}^{a+1,b^{a+1}_{max}}\}$. 
$\{\mathcal{I}_{tem}^{a+1,0},\dots, \mathcal{I}_{tem}^{a+1,b^{a+1}_{max}}\}$= DotSrch $(G,\mathcal{I}_k^{a,b},\mathcal{I})$ }
\label{tab:search $dot$}
\begin{algorithmic}[1]
\State  \textbf{Initializations:} $\mathcal{I}_{tem}^{a+1,b} \gets \emptyset, ~ \forall~b \leq b^{a+1}_{max}$.
\For{each instance of LETS structure $\mathcal{S}$ in $\mathcal{I}_k^{a,b}$}
\parState {Consider $\mathcal{V}$ to be the set of variable nodes in $V \setminus \mathcal{S}$, which have at least two connections with the check nodes in $\Gamma_{o}{(\mathcal{S})}$ and have no connection with the check nodes in $\Gamma_{e}{(\mathcal{S})}$. }
\vspace{-6pt}
\For {each variable node $v \in \mathcal{V}$}
\State $\mathcal{S}'=\mathcal{S} \cup v$.
\If{$|\Gamma_{o}{(\mathcal{S}')}| \leq b^{a+1}_{max}$}
\State $\mathcal{I}_{tem}^{a+1, |\Gamma_{o}{(\mathcal{S}')}|}=\mathcal{I}_{tem}^{a+1, |\Gamma_{o}{(\mathcal{S}')}|} \cup  \{ \mathcal{S}'\setminus \mathcal{I}\}$.
\EndIf
\EndFor
\EndFor
\State \textbf{Output:} $\{\mathcal{I}_{tem}^{a+1,0},\dots, \mathcal{I}_{tem}^{a+1,b^{a+1}_{max}}\}$.
\end{algorithmic}
\end{routine}
\begin{routine}
\centering
\caption{{\bf (PathSrch)} Expanding all the instances $\mathcal{I}_k^{a,b}$ of LETS structures in the $(a,b)$ class of Tanner graph $G$ using $pa_m$ to find instances of LETS structures of size $a+m$, excluding the already found instances $\mathcal{I}$, and storing the rest in $\{\mathcal{I}_{tem}^{a+m,0},\dots, \mathcal{I}_{tem}^{a+m,b^{a+m}_{max}}\}$.  
$\{\mathcal{I}_{tem}^{a+m,0},\dots, \mathcal{I}_{tem}^{a+m,b^{a+m}_{max}}\}$= PathSrch$(G,\mathcal{I}_k^{a,b}, \mathcal{I}, m)$}
\label{tab:search $path$}
\begin{algorithmic}[1]
\State  \textbf{Initializations:} $\mathcal{I}_{tem}^{a+m,b} \gets \emptyset, ~ \forall~b \leq b^{a+m}_{max}$.
\parFor{each LETS instance ${\cal S}$ in $\mathcal{I}_k^{a,b}$}
\parFor {each unsatisfied check node $c_k \in \Gamma_{o}{(\mathcal{S})}$}
\parState {Find all the paths $\mathcal{PA}_k$ of length $m$ in the Tanner graph, starting from $c_k$. }
\EndparFor
\parFor {any pair of unsatisfied check nodes $c_k$ and $c_j$ in $\Gamma_{o}{(\mathcal{S})}$, where $k \neq j$}
\parFor {any path $pa \in \mathcal{PA}_k$ and any path $pa' \in \mathcal{PA}_j$}
\parIf{$pa$ and $pa'$ end with the same node and this node is their only common node, and if variable and check nodes of $pa$ and $pa'$ are not in $G(\mathcal{S})$}
\parState {$\mathcal{S}'=\mathcal{S}~\cup$ \{set of variable nodes in $pa \cup pa'$\}.}
\If{$|\Gamma_{o}{(\mathcal{S}')}| \leq b^{a+m}_{max}$}
\State $\mathcal{I}_{tem}^{a+m, |\Gamma_{o}{(\mathcal{S}')}|}=\mathcal{I}_{tem}^{a+m, |\Gamma_{o}{(\mathcal{S}')}|} \cup  \{ \mathcal{S}'\setminus \mathcal{I}\}$.
\EndIf
\EndparIf
\EndparFor
\EndparFor
\EndparFor
\State \textbf{Outputs:} $\{\mathcal{I}_{tem}^{a+m,0},\dots, \mathcal{I}_{tem}^{a+m,b^{a+m}_{max}}\}$.
\end{algorithmic}
\end{routine}
\begin{routine}
\centering
\caption{{\bf (LolliSrch)} Expanding all the instances $\mathcal{I}_k^{a,b}$ of LETS structures in the $(a,b)$ class of Tanner graph $G$ using $lo^c_m$ to find instances of LETS structures of size $a+m$, excluding the already found instances $\mathcal{I}$, and storing the rest in $\{\mathcal{I}_{tem}^{a+m,0},\dots, \mathcal{I}_{tem}^{a+m,b^{a+m}_{max}}\}$. The set of instances of simple cycles of length c, $\mathcal{I}_c$, is also
an input. 
 $\{\mathcal{I}_{tem}^{a+m,0},\dots, \mathcal{I}_{tem}^{a+m,b^{a+m}_{max}}\}$= LolliSrch $(G,\mathcal{I}_k^{a,b},\mathcal{I},\mathcal{I}_c, m)$}
\label{tab:search lolli}
\begin{algorithmic}[1]
\State  \textbf{Initializations:} $\mathcal{I}_{tem}^{a+m,b} \gets \emptyset ,~ \forall~b \leq b^{a+m}_{max}$,  $d=m+1-c$.
\parFor{each LETS instance ${\cal S}$ in $\mathcal{I}_k^{a,b}$}
\parState {Find all the paths ${\cal PA}$ of length $2(d-1)$ in the Tanner graph, starting from check nodes $c'$ in $\Gamma_{o}{(\mathcal{S})}$ that have no common nodes with $G({\cal S})$ other than $c'$. }
\parFor {each structure $\mathcal{C} \in \mathcal{I}_c$, for which $G({\cal C})$ has no common node with $G({\cal S})$, let $\Gamma_{o}(\mathcal{C})$ denote the set of unsatisfied check nodes of $G(\mathcal{C})$}
\vspace{-6pt}
\For {each path, $pa \in \mathcal{PA}$} \label{lolli-comp}
\parIf {$pa$ ends with a check node $c''$ in $\Gamma_{o}(\mathcal{C})$ and if $c''$ is the only common node between $pa$ and $\Gamma_{o}(\mathcal{C})$}
\parState {$\mathcal{S}'= \{\mathcal{S}~\cup$ \{set of variable nodes in $pa \cup G(\mathcal{C})$\}.}
\If{$|\Gamma_{o}{(\mathcal{S}')}| \leq b^{a+m}_{max}$}
\State $\mathcal{I}_{tem}^{a+m, |\Gamma_{o}{(\mathcal{S}')}|}=\mathcal{I}_{tem}^{a+m, |\Gamma_{o}{(\mathcal{S}')}|} \cup  \{ \mathcal{S}'\setminus \mathcal{I}\}$.
\EndIf
\EndparIf
\EndFor
\EndparFor
\EndparFor
\State \textbf{Output:} $\{\mathcal{I}_{tem}^{a+m,0},\dots, \mathcal{I}_{tem}^{a+m,b^{a+m}_{max}}\}$.
\end{algorithmic}
\end{routine}
The Routines are similar to those of~\cite{hashemireg} for variable-regular graphs, with the difference being that, due to the presence of variable nodes with different degrees, after 
applying an expansion technique, the algorithm needs to check whether the resultant LETS instances are in the interest range.

The complexity of the search algorithm depends, in general, on the multiplicity of different
instances of LETS structures and the expansion techniques used in different classes. A detailed discussion on the complexity of the dpl-based search algorithm for variable-regular graphs can 
be found in \cite{hashemireg}. The generalization of those discussions to irregular graphs is rather simple and thus not presented here.

\section{Efficient Exhaustive Search of Elementary Trapping sets for Irregular LDPC Codes}
\label{sec:ETS}

\begin{figure}[] 
\centering
\includegraphics [width=0.4\textwidth]{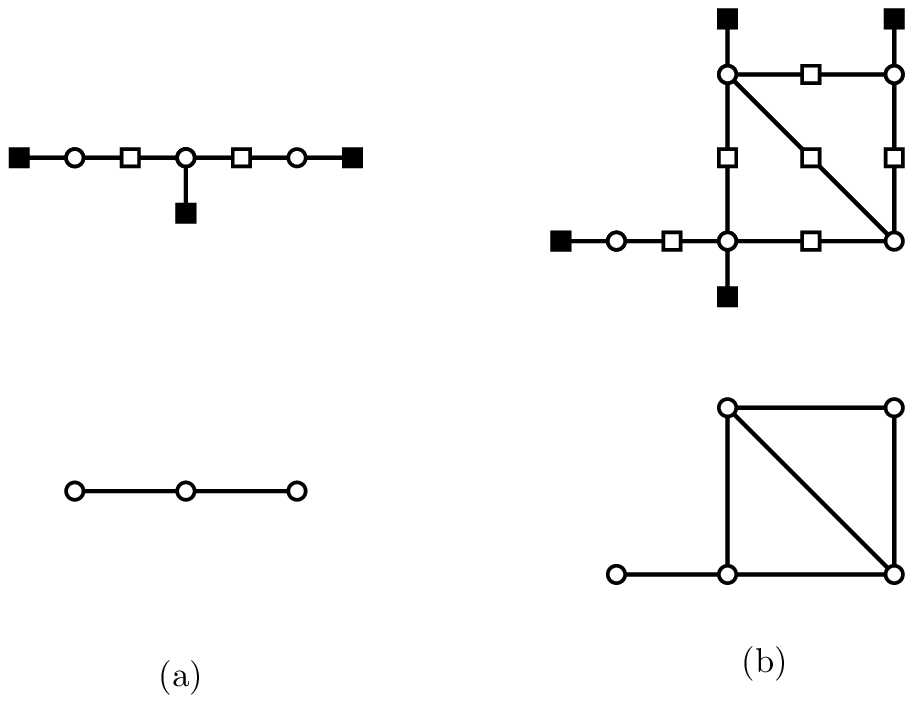}
\caption{Two ETSLs in the (a) $(3,3)$ and (b) $(5,4)$ classes, respectively.}
\label{fig:ETSCgen}
\end{figure}

Leafless ETSs (LETSs) are known to be the main problematic structures in the error floor region of variable-regular LDPC codes. For irregular LDPC codes, however, in addition to LETSs, there are other ETSs that are problematic but are not leafless, 
i.e., they have variable nodes that are connected to only one satisfied check node~ \cite{milenkovic2007asymptotic}, \cite{abu2010trapping},\cite{mehdi2012}, \cite{kyung2012finding}, \cite{sinathesis}, \cite{Fals}. 
This is particularly the case for irregular codes with degree-2 variable nodes~\cite{mehdi2012},~\cite{kyung2012finding}. 
In Sections \ref{sec:char/search} and \ref{sec:search}, we studied the LETS structures of irregular LDPC codes. This section is dedicated to ETSs of irregular codes that have leaves. We use the notation ``ETSL'' for such trapping sets, and remind the reader that it is the the normal graph representation of ETSL structures that has at least one leaf. 
Two examples of ETSL structures in the $(3,3)$ and $(5,4)$ classes, along with their normal graphs, are shown in Fig. \ref{fig:ETSCgen}. 

\begin{figure}[] 
\centering
\includegraphics [width=0.2\textwidth]{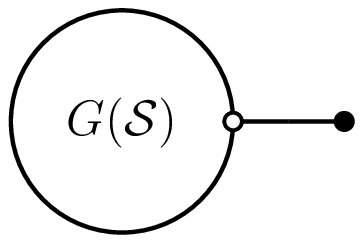}
\caption{Expansion of an ETS structure $\mathcal{S}$ with  $dot_1^k$.}
\label{fig:dtt}
\end{figure}

The \textit{depth-one tree (dot)} expansion technique plays an important role in the characterization and search of ETSLs. However, unlike the LETS case, where $dot_m^k$ expansion with $m \geq 2$ was used, in the ETSL case, we
are interested in the $dot_m^k$ expansion with $m=1$. 
Fig. \ref{fig:dtt} shows a structure expanded by $dot_1^k$.

\begin{lem}
\label{lem:cha}
Suppose that $\mathcal{S}$ is an $(a,b)$ ETS structure of irregular Tanner graphs.
Expansion of $\mathcal{S}$ using $dot^k_1$, results in 
ETS structure(s) in the  $(a+1,b+k-2)$ class, where $k$ is the degree of the new variable node added to $\mathcal{S}$. 
\end{lem}
 
In general, the ETSL structures can be partitioned into two categories. The ETSLs that contain at least one LETS sub-structure, and those that do not contain any LETS sub-structure. 
We use notations {ETSL}$_1$ and {ETSL}$_2$, to represent these two categories, respectively.
The structures in Fig. \ref{fig:ETSCgen} (a) and (b) are examples of {ETSL}$_2$ and {ETSL}$_1$, respectively. In the rest of this section,  {ETSL}$_1$ and {ETSL}$_2$ 
structures are characterized and an efficient search algorithm is presented to find them in an exhaustive fashion. 

\subsection{Characterization of {ETSL}$_1$}
\label{subsecfr}
We characterize {ETSL}$_1$ structures as the expansions of their LETS sub-structures through the following proposition. 

\begin{pro}
\label{ETSL2}
Any $(a,b)$ {ETSL}$_1$ structure ${\cal S}$ of irregular graphs, with minimum variable node degree $d_{v_{min}} \geq 2$,  in the range of $a \leq a_{max}$ and $b \leq b_{max}$, can be obtained by successive application of $dot^k_1$ to the 
largest $(a',b')$ LETS substructure of ${\cal S}$, where $a' < a_{max}$ and $b' \leq b_{max}$.  
\end{pro}

\begin{proof}
Suppose that $\mathcal{S}=(F,E)$, 
and that $\mathcal{S}'=(F',E')$ is the largest LETS sub-structure of $\mathcal{S}$. We assume that the expansion $F \backslash F'$ is connected. 
Disconnected expansions can be treated as a collection of connected expansions. We first prove that the number of edges connecting $F \backslash F'$ to the nodes in $\mathcal{S}'$ is one. 
Suppose that the number of such edges is at least 2. If at least two of these edges are connected to one node, $v$, in $F \backslash F'$, then adding $v$  to $\mathcal{S}'$ results in a LETS
sub-structure of $\mathcal{S}$ that is larger than $\mathcal{S}'$. This is a contradiction. So the connecting edges must be each connected to a distinct node in $F \backslash F'$. Consider two such nodes and call them 
$v_1$ and $v_2$. Since the expansion is connected, there is at least one path between $v_1$ and $v_2$ in the expansion. One can thus add all the nodes on that path plus $v_1$ and $v_2$ to $\mathcal{S}'$, and obtain a LETS substructure of $\mathcal{S}$ larger than $\mathcal{S}'$. Again, a contradiction. We thus conclude that the expansion is connected by only one edge to $\mathcal{S}'$. (Suppose that the node in $F \backslash F'$ which is connected to $\mathcal{S}'$ is $v$.)
With a similar approach, it can be shown that the expansion can not contain a cycle. Since the expansion has only one connection to $\mathcal{S}'$ and does not contain a cycle, the only possibility is a rooted tree with the root at node $v$. It is now easy to see that a rooted tree, as an expansion, can be implemented by successive application of $dot^k_1$ expansions. Using Lemma~\ref{lem:cha}, one can see that as $dot^k_1$ expansions with $k \geq 2$ are applied to the LETS structure $\mathcal{S}'$, by each application, the $a$ value is increased by one, and the $b$ value is either increased or remains the same. This implies that if for the target  $(a,b)$ {ETSL}$_1$ structure ${\cal S}$, we have $a \leq a_{max}$ and $b \leq b_{max}$, then for the largest $(a',b')$ LETS substructure of ${\cal S}$, we must have $a' < a_{max}$ and $b' \leq b_{max}$.
\end{proof}

Based on Proposition~\ref{ETSL2}, to find all the  ETSL$_1$s of an irregular LDPC code in the interest range of $a \leq a_{max}$ and $b \leq b_{max}$, 
one can first find all the instances of  $(a,b)$ LETS structures with $a \leq a_{max}$ and $b \leq b_{max}$, and then apply $dot^k_1$ expansions successively to these instances, 
as appropriate. One should note that all the instances generated in the process of the application of $dot^k_1$ expansions to a LETS are ETSL$_1$s. 
The following corollary determines which $dot^k_1$ expansion should be applied to which LETS  or {ETSL}$_1$ instance.

 \begin{cor}
 \label{corhb}
 To generate all the $(a,b)$ ETSL$_1$ instances of an irregular graph with $d_{v_{min}} \geq 2$, in the range of  $a \leq a_{max}$ and $b \leq b_{max}$, for each
 value of $i$ in the range $d_{v_{min}}-2 \leq i \leq b_{max}-1$, the expansion $dot^k_1$ with $k \leq i+2$ should be applied to 
 any LETS and  {ETSL}$_1$s with $a \leq a_{max}-1$ and $b= b_{max}-i$.
 \end{cor}

\begin{ex}
 Based on Corollary~\ref{corhb}, if one is interested in finding all {ETSL}$_1$s  in the interest range of $a \leq a_{max}$ and $b \leq 2$, for an irregular graph with $d_{v_{min}}=2$, 
 expansion $dot^2_1$ should be applied to LETSs and  {ETSL}$_1$s with $a \leq a_{max}-1$ and $b=2$, and $dot^2_1$ and $dot^3_1$ expansions should be applied to LETSs and  {ETSL}$_1$s 
 with $a \leq a_{max}-1$ and $b=1$.
\end{ex}

\begin{rem} 
Although having variable nodes of degree one is not common in most LDPC codes, the results presented here can easily be generalized to the case where the code has such variable nodes. 
\end{rem}

\subsection{Characterization of {ETSL}$_2$}
\label{subsecvg}
It is easy to see that ETSL$_2$ structures (in the space of normal graphs) contain no cycles, and are thus trees. The following proposition is then simple to prove.

\begin{pro}
\label{proks}
Any $(a,b)$ {ETSL}$_2$ structure of irregular graphs with $d_{v_{min}} \geq 2$, in the range $a \leq a_{max}$ and $b \leq b_{max}$, can be obtained by successive application of $dot^k_1$ expansions to single variable nodes with degree less than or equal to $b_{max}$.
\end{pro}
 
Based on Proposition~\ref{proks}, for the special case of $b_{max}=2$, the only configuration of {ETSL}$_2$ is a chain consisting of only degree-$2$ variable nodes. 

\subsection{Exhaustive Search of ETSLs in Irregular Tanner Graphs}
\label{subsecfe}

Based on the results of Subsections~\ref{subsecfr} and \ref{subsecvg}, Algorithm \ref{algnon} provides a pseudo code for finding all the instances of all the $(a,b)$ {ETSL} structures of an irregular Tanner graph 
in the interest range of $a \leq a_{max}$ and $b \leq b_{max}$, exhaustively.
\begin{algorithm}
\caption{{\bf (ETSL Exhaustive Search)} Finds all the instances of $(a,b)$ ETSL structures of an irregular Tanner graph $G=(V,E)$ with girth $g$ and $d_{v_{min}} \geq 2$, for $a \leq a_{max}$ and $b \leq b_{max}$. 
The input is all the instances of $(a,b)$ LETS structures, in the range $a \leq a_{max}$ and $b \leq b_{max}$, $\mathcal{I}$. ($\mathcal{I}^{a,b}$ is the set of all the instances in the $(a,b)$ class.) The output is the set ${\cal I}_{ETSL}$, which contains all the instances 
of {ETSL}$_1$s and {ETSL}$_2$s in the interest range.}
\label{algnon}
 \begin{algorithmic} [1] 
\State \textbf{Inputs:} $G, a_{max}, \mathcal{I}, g$.
\State  \textbf{Initializations:}  ${\cal I}_{tem}^{a,b} \gets \emptyset,  {\cal I}_{ETSL}^{a,b} \gets \emptyset,~\forall ~a \leq a_{max}, ~\forall ~b \leq b_{max}$.
\For{$b=2, \ldots, b_{max}$}
\State {${\cal I}_{tem}^{1,b}$ = the set of variable nodes of degree $b$ in $G$.}
\EndFor
\For {$a=1, \dots, a_{max}-1$}
\For{$b=1, \ldots, b_{max}$}
\State ${\cal I}_{tem}^{a,b}={\cal I}_{tem}^{a,b} \cup {\cal I}^{a,b}$.
\For {any structure $\mathcal{S} \in {\cal I}_{tem}^{a,b}$}
\parState {Consider $\mathcal{V}$ to be the set of variable nodes in $V \setminus \mathcal{S}$ with degrees less than or equal to $b_{max}+2-b$, which have only one connection to the check nodes in $\Gamma_{o}{(\mathcal{S})}$ and have no connection to the check nodes in $\Gamma_{e}{(\mathcal{S})}$,}
\vspace{1pt}
\For {each variable node $v \in \mathcal{V}$}
\parState {${\cal I}_{ETSL}^{a+1,b'}= {\cal I}_{ETSL}^{a+1,b'} \cup \{\mathcal{S} \cup v\}$, ${\cal I}_{tem}^{a+1,b'}={\cal I}_{tem}^{a+1,b'} \cup \{\mathcal{S} \cup v\}$, where $b'=b+deg(v)-2$.}
\EndFor
\EndFor
\EndFor
\EndFor
\State \textbf{Output:} ${\cal I}_{ETSL}= \{{\cal I}_{ETSL}^{a,b},~\forall ~a \leq a_{max}, ~\forall ~b \leq b_{max}\}$.
\end{algorithmic}
\end{algorithm}

We note that the complexity of finding ETSLs is often negligible (less than 1\% in our experiments) in comparison with the complexity of search for LETSs. 

\section{Numerical Results}
\label{sec:numerical}

We have applied the proposed search algorithm of Section~\ref{sec:search} (for LETSs) and that of Section~\ref{sec:ETS} (for ETSL) to find the ETSs of a large 
number of irregular LDPC codes with a wide range of variable node degrees, rates and block lengths, exhaustively. These codes and their parameters including block length, rate, girth and variable 
and check node degree distributions are listed in Table \ref{tab:codelist}. 
For each code, we have used both the upper bounds of~(\ref{eqkd}) and (\ref{eqif}), and observed that, although we have no proof that the latter will result in an exhaustive search in general, for all the codes 
tested in this work, both upper bounds provide an exhaustive coverage of LETSs.

\begin{table*}[]
\centering
\caption{List of irregular LDPC Codes Used in This Paper}
\setlength{\tabcolsep}{1pt}
\label{tab:codelist}
\begin{tabular}{||c|c|c|c|c|c|c|| }
\cline{1-7}
Code &$n$&$R$&$g$&$\lambda(x)$&$\rho(x)$&
Ref\\
\cline{1-7}
\multirow{2}{*}{$\mathcal{C}_{1}$}&\multirow{2}{*}{1000}&\multirow{2}{*}{0.7}&\multirow{2}{*}{6}&\multirow{2}{*}{$\lambda(x)=0.243x^2+0.757x^3$}&$\rho(x) =0.002x^7+0.012x^8+0.043x^9+0.151x^{10}+$&\multirow{10}{*}{\cite{radford}}\\
&&&&&$0.318x^{11}+0.264x^{12}+0.144x^{13}+0.052x^{14}+0.014x^{15}$&\\
\cline{1-6}
\multirow{2}{*}{$\mathcal{C}_{2}$}&\multirow{2}{*}{8000}&\multirow{2}{*}{0.78}&\multirow{2}{*}{6}&\multirow{2}{*}{$\lambda(x)=0.636x^2+0.364x^3$}&$\rho(x) =0.001x^{11}+0.020x^{12}+0.345x^{13}$&\\
&&&&&$+0.557x^{14}+0.068x^{15}+0.008x^{16}+0.001x^{17}$&\\
\cline{1-6}
$\mathcal{C}_{3}$&4000&0.5&6&$\lambda(x) =0.871x^2+0.129x^3$&$\rho(x) =0.008x^4+0.757x^5+0.231x^6+0.004x^7$&\\
\cline{1-6}
$\mathcal{C}_{4}$&500&0.5&6&$\lambda(x) =0.636x^2+0.364x^3$&$\rho(x) =0.030x^4+0.360x^5+0.530x^6+0.063x^7+0.017x^8$&\\
\cline{1-6}
\multirow{2}{*}{$\mathcal{C}_{5}$}&\multirow{2}{*}{1000}&\multirow{2}{*}{0.5}&\multirow{2}{*}{6}&\multirow{2}{*}{$\lambda(x)=0.762x^3+0.238x^4$}&$\rho(x) =0.008x^{5}+0.062x^{6}+0.480x^{7}$&\\
&&&&&$+0.358x^{8}+0.079x^{9}+0.01x^{10}+0.003x^{11}$&\\
\cline{1-6}
\multirow{2}{*}{$\mathcal{C}_{6}$}&\multirow{2}{*}{300}&\multirow{2}{*}{0.5}&\multirow{2}{*}{6}&\multirow{2}{*}{$\lambda(x) =0.878x^3+0.122x^4$}&$\rho(x) =0.019x^3+0.110x^4+0.522x^5$&\\
&&&&&$+0.302x^6+0.039x^7+0.008x^9$&\\
\cline{1-7}
$\mathcal{C}_{7}$&576&0.75&6&\multirow{3}{*}{$\lambda(x) =0.114x+0.409x^2+0.477x^5$}&\multirow{3}{*}{$\rho(x) =0.318x^{13}+0.682x^{14}$}&\multirow{9}{*}{\cite{802.16}}\\
\cline{1-4}
$\mathcal{C}_{8}$&1056&0.75&6&&&\\
\cline{1-4}
$\mathcal{C}_{9}$&2304&0.75&6&&&\\
\cline{1-6}
$\mathcal{C}_{10}$&576&0.66&6&\multirow{3}{*}{$\lambda(x) =0.173x+0.037x^2+0.790x^3$}&\multirow{3}{*}{$\rho(x) =0.864x^9+0.136x^{10}$}&\\
\cline{1-4}
$\mathcal{C}_{11}$&1056&0.66&6&&&\\
\cline{1-4}
$\mathcal{C}_{12}$&2304&0.66&6&&&\\
\cline{1-6}
$\mathcal{C}_{13}$&576&0.5&6&\multirow{3}{*}{$\lambda(x) =0.289x+0.316x^2+0.395x^5$}&\multirow{3}{*}{$\rho(x) =0.632x^5+0.368x^6$}&\\
\cline{1-4}
$\mathcal{C}_{14}$&1056&0.5&6&&&\\
\cline{1-4}
$\mathcal{C}_{15}$&2304&0.5&6&&&\\
\cline{1-7}
\multirow{2}{*}{$\mathcal{C}_{16}$}&\multirow{2}{*}{1944}&\multirow{2}{*}{0.5}&\multirow{2}{*}{6}&$\lambda(x) =0.256x+0.314x^2$&\multirow{2}{*}{$\rho(x) =0.814x^6+0.186x^7$}&\multirow{2}{*}{\cite{802.11}}\\
&&&&$+0.046x^3+0.384x^{10}$&&\\
\cline{1-7}
\multirow{2}{*}{$\mathcal{C}_{17}$}&\multirow{2}{*}{504}&\multirow{2}{*}{0.5}&\multirow{2}{*}{8}&$\lambda(x) =0.239x+0.210x^2+0.036x^3+$&\multirow{2}{*}{$\rho(x) =0.077x^6+0.834x^7+0.089x^8$}&\multirow{4}{*}{\cite{mackayencyclopedia}}\\
&&&&$0.122x^4+0.014x^6+0.007x^{13}+0.372x^{14}$&&\\
\cline{1-6}
\multirow{2}{*}{$\mathcal{C}_{18}$}&\multirow{2}{*}{1008}&\multirow{2}{*}{0.5}&\multirow{2}{*}{8}&$\lambda(x) =0.239x+0.210x^2+0.035x^3+$&\multirow{2}{*}{$\rho(x) =0.009x^6+0.978x^7+0.013x^8$}&\\
&&&&$0.121x^4+0.016x^6+0.003x^{13}+0.376x^{14}$&&\\
\cline{1-7}
\end{tabular}
\end{table*}

For all the run-times reported in this paper, a desktop computer with $2.4$-GHz CPU and $8$-GB RAM is used, and the search algorithms are implemented in MATLAB.
Except codes $\mathcal{C}_1$-$\mathcal{C}_6$, $\mathcal{C}_{17}$, and $\mathcal{C}_{18}$, the other LDPC codes are all structured codes.
Codes $\mathcal{C}_{7}$-$\mathcal{C}_{15}$ are the LDPC codes used in the IEEE 802.16e standard~\cite{802.16}, and Code $\mathcal{C}_{16}$ is a code used in the IEEE 802.11n~\cite{802.11}.
For structured codes, their structural properties are used to simplify the search.

Tables \ref{tab:341}-\ref{tab:peg1} list the multiplicity of instances of ETSs, LETSs, EASs and FEASs in different $(a,b)$ classes, $a \leq a_{max}, b \leq b_{max}$, for these codes.
Each row of a table corresponds to a non-empty ETS class, and for each class, the total number of instances of ETSs, LETSs, EASs and FEASs are listed. The difference between the total number of ETSs and LETSs gives the total number of ETSLs.
In the last two rows of each table, the run-times of the search algorithm based on the upper bounds of (\ref{eqkd}) and (\ref{eqif}) are reported, respectively. 
Comparison of the run-times shows that large improvements in the search speed, in some cases more than an order of magnitude, can be obtained  by using (\ref{eqif}) instead of (\ref{eqkd}).
We also note that, for each code, only less than 1\% of the reported run-time is for finding ETSLs. For example, while most of the ETSs of $C_{15}$, reported in Table \ref{tab:50}, are ETSLs, 
it takes Algorithm \ref{algnon} only $2$ seconds to find all the ETSLs of this code.

\begin{table}[]
\centering
\caption{Multiplicities of $(a,b)$ ETSs, LETSs, EASs and FEASs of Codes $\mathcal{C}_{1}$ and $\mathcal{C}_{2}$ within the range of $a \leq 7$ and $b \leq 3$}
\setlength{\tabcolsep}{1pt}
\label{tab:341}
\begin{tabular}{||c|c|c|c|c|||c|c|c|c|| }
\cline{1-9}
&\multicolumn{4}{c|||}{$\mathcal{C}_1$} &\multicolumn{4}{c||}{$\mathcal{C}_2$}\\
\cline{1-9}
$(a,b)$&Total&Total&Total&Total&Total&Total&Total&Total\\
class&ETS&LETS&EAS&FEAS&ETS&LETS&EAS&FEAS\\
\cline{1-9}
(3,3)&32&32&32&23&904&904&904&854\\
\cline{1-9}
(4,2)&1&1&1&1&13&13&13&13\\
\cline{1-9}
(4,3)&19&19&10&8&52&52&27&25\\
\cline{1-9}
(5,2)&0&0&0&0&2&2&2&2\\
\cline{1-9}
(5,3)&28&27&23&21&758&540&540&508\\
\cline{1-9}
(6,2)&1&1&1&1&13&13&13&12\\
\cline{1-9}
(6,3)&34&34&30&18&127&93&77&70\\
\cline{1-9}
(7,1)&0&0&0&0&1&1&1&1\\
\cline{1-9}
(7,2)&1&1&1&1&4&4&4&4\\
\cline{1-9}
(7,3)&79&70&56&43&777&541&538&506\\
\cline{1-9}
Search w. (\ref{eqkd})&\multicolumn{4}{c|||}{33 min.} &\multicolumn{4}{c||}{36 min.}\\
\cline{1-9}
Search w. (\ref{eqif})&\multicolumn{4}{c|||}{13 min.} &\multicolumn{4}{c||}{29 min.}\\
\cline{1-9}
\end{tabular}
\end{table}

To compare the complexity of the two search techniques discussed in Sections~\ref{sec:char/search} and \ref{sec:search}, we consider the two examples of $\mathcal{C}_1$ and $\mathcal{C}_{16}$.
For $\mathcal{C}_1$, since the variable degrees  are rather small ($3$ and $4$), and we are interested in ETSs with relatively small values of $a$ and $b$ ($a \leq 7$ and $b \leq 3$), the search algorithm proposed 
in Section \ref{sec:char/search}, based on the characterization Table \ref{tab:4,6c3}, is quite efficient and can find all the ETS structures reported in Table \ref{tab:341} in only 21 minutes (compared to $33$ minutes for Algorithm~\ref{alg2} with (\ref{eqkd})). For $\mathcal{C}_{16}$, however, the algorithm of Section \ref{sec:char/search} becomes clearly inefficient, since based on Theorem \ref{theo:main}, one should use the characterization table of a variable-regular graph 
with $d_v=11$, in the range of $a\leq 12$ and $b \leq 110$! The inefficiency of the algorithm in this case is due to the wide range of variable degrees for this code ($2,3,4,11$), and the relatively large range of $a$ values ($a \leq 12$).

\begin{table}[]
\centering
\caption{Multiplicities of  $(a,b)$ ETSs, LETSs, ETSs and FEASs of Codes $\mathcal{C}_{3}$ and $\mathcal{C}_{4}$ within the range of $a \leq 9$ and $b \leq 4$}
\setlength{\tabcolsep}{1pt}
\label{tab:34}
\begin{tabular}{||c|c|c|c|c|||c|c|c|c|| }
\cline{1-9}
 &\multicolumn{4}{c|||}{$\mathcal{C}_3$}&\multicolumn{4}{c||}{$\mathcal{C}_4$}\\
\cline{1-9}
$(a,b)$&Total&Total&Total&Total&Total&Total&Total&Total\\
class&ETS&LETS&EAS&FEAS&ETS&LETS&EAS&FEAS\\
\cline{1-9}
(2,4)&24610&0&0&0&1938&0&0&0\\
\cline{1-9}
(3,3)&110&110&110&110&62&62&62&58\\
\cline{1-9}
(3,4)&96&96&0&0&169&169&0&0\\
\cline{1-9}
(4,3)&1&1&1&1&13&13&9&9\\
\cline{1-9}
(4,4)&2336&826&826&804&1077&397&386&303\\
\cline{1-9}
(5,3)&13&13&13&11&42&42&42&37\\
\cline{1-9}
(5,4)&38&26&16&15&375&243&161&126\\
\cline{1-9}
(6,2)&1&1&1&1&1&1&1&1\\
\cline{1-9}
(6,3)&1&1&1&1&18&18&15&14\\
\cline{1-9}
(6,4)&409&231&231&221&1145&680&641&502\\
\cline{1-9}
(7,2)&0&0&0&0&1&1&1&1\\
\cline{1-9}
(7,3)&15&6&6&6&54&46&44&37\\
\cline{1-9}
(7,4)&25&10&8&8&762&579&473&357\\
\cline{1-9}
(8,2)&0&0&0&0&4&4&4&3\\
\cline{1-9}
(8,3)&0&0&0&0&47&39&39&35\\
\cline{1-9}
(8,4)&249&72&72&70&1831&1283&1203&918\\
\cline{1-9}
(9,1)&0&0&0&0&1&1&1&1\\
\cline{1-9}
(9,3)&3&3&3&2&110&81&78&62\\
\cline{1-9}
(9,4)&7&7&5&5&2001&1521&1331&1015\\
\cline{1-9}
Search w. (\ref{eqkd}) &\multicolumn{4}{c|||}{16 min.}&\multicolumn{4}{c||}{57 min.}\\
\cline{1-9}
Search w. (\ref{eqif}) &\multicolumn{4}{c|||}{15 min.}&\multicolumn{4}{c||}{55 min.}\\
\cline{1-9}
\end{tabular}
\end{table}

\begin{table}[]
\centering
\caption{Multiplicities of  $(a,b)$ ETSs, LETSs, EASs and FEASs of Codes $\mathcal{C}_{5}$ and $\mathcal{C}_{6}$ within the range of $a \leq 8$ and $b \leq 7$}
\setlength{\tabcolsep}{1pt}
\label{tab:45}
\begin{tabular}{||c|c|c|c|c|||c|c|c|c|| } 
\cline{1-9}
&\multicolumn{4}{c|||}{$\mathcal{C}_5$}&\multicolumn{4}{c||}{$\mathcal{C}_6$}\\
\cline{1-9}
$(a,b)$&Total&Total&Total&Total&Total&Total&Total&Total\\
class&ETS&LETS&EAS&FEAS&ETS&LETS&EAS&FEAS\\
\cline{1-9}
(2,6)&9146&0&0&0&2516&0&0&0\\
\cline{1-9}
(2,7)&5673&0&0&0&672&0&0&0\\
\cline{1-9}
(3,6)&830&830&0&0&486&486&0&0\\
\cline{1-9}
(3,7)&1034&1034&0&0&256&256&0&0\\
\cline{1-9}
(4,4)&0&0&0&0&1&1&1&1\\
\cline{1-9}
(4,6)&150&150&0&0&180&180&0&0\\
\cline{1-9}
(4,7)&275&275&0&0&168&168&0&0\\
\cline{1-9}
(5,6)&35&35&1&1&110&93&0&0\\
\cline{1-9}
(5,7)&121&121&0&0&149&146&0&0\\
\cline{1-9}
(6,6)&15&15&1&1&69&69&8&4\\
\cline{1-9}
(6,7)&52&52&2&0&131&131&5&1\\
\cline{1-9}
(7,6)&5&5&0&0&49&49&6&2\\
\cline{1-9}
(7,7)&23&23&3&1&133&133&14&1\\
\cline{1-9}
(8,5)&0&0&0&0&2&2&2&1\\
\cline{1-9}
(8,6)&0&0&0&0&40&40&9&3\\
\cline{1-9}
(8,7)&13&13&2&1&151&151&20&7\\
\cline{1-9}
Search w. (\ref{eqkd}) &\multicolumn{4}{c|||}{219 min.}&\multicolumn{4}{c||}{42 min.}\\
\cline{1-9}
Search w. (\ref{eqif}) &\multicolumn{4}{c|||}{15 min.}&\multicolumn{4}{c||}{6 min.}\\
\cline{1-9}
\end{tabular}
\end{table}

\begin{table*}[]
\centering
\caption{Multiplicities of  $(a,b)$ ETSs, LETSs, EASs and FEASs of Codes  $\mathcal{C}_{7}$-$\mathcal{C}_{12}$  within the range of $a \leq 8$ and $b \leq 2$}
\setlength{\tabcolsep}{.5pt}
\label{tab:75}
\begin{tabular}{||c|c|c|c|c|||c|c|c|c|||c|c|c|c|| } 
\cline{1-13}
&\multicolumn{4}{c|||}{$\mathcal{C}_{7}$} &\multicolumn{4}{c|||}{$\mathcal{C}_{8}$}&\multicolumn{4}{c||}{$\mathcal{C}_{9}$}\\
\cline{1-13}
$(a,b)$&Total&Total&Total&Total&Total&Total
&Total&Total&Total&Total&Total&Total\\
class&ETS&LETS&EAS&FEAS&ETS&LETS&EAS&FEAS&ETS&LETS&EAS&FEAS\\
\cline{1-13}
(2,2)&96&0&0&0&176&0&0&0&384&0&0&0\\
\cline{1-13}
(3,2)&72&0&0&0&132&0&0&0&288&0&0&0\\
\cline{1-13}
(4,2)&192&144&144&0&176&88&88&0&192&0&0&0\\
\cline{1-13}
(5,2)&720&216&216&0&616&308&308&0&384&288&288&0\\
\cline{1-13}
(6,1)&48&48&48&0&44&44&44&0&96&96&96&0\\
\cline{1-13}
(6,2)&2556&1068&1068&0&1958&418&418&0&1200&144&144&0\\
\cline{1-13}
(7,1)&336&240&240&0&176&88&88&0&192&0&0&0\\
\cline{1-13}
(7,2)&9264&3600&3600&0&5192&1144&1144&0&2880&672&672&0\\
\cline{1-13}
(8,0)&48&48&48&48&0&0&0&0&0&0&0&0\\
\cline{1-13}
(8,1)&1176&720&720&0&440&220&220&0&288&96&96&0\\
\cline{1-13}
(8,2)&35040&13464&13464&0&16104&5368&5368&0&10176&2304&2304&0\\
\cline{1-13}
Search w. (\ref{eqkd}) &\multicolumn{4}{c|||}{301 min.}&\multicolumn{4}{c|||}{115 min.}&\multicolumn{4}{c||}{43 min.}\\
\cline{1-13}
Search w. (\ref{eqif}) &\multicolumn{4}{c|||}{69 min.}&\multicolumn{4}{c|||}{31 min.}&\multicolumn{4}{c||}{13 min.}\\
\cline{1-13}
\hline
\hline
&\multicolumn{4}{c|||}{$\mathcal{C}_{10}$} &\multicolumn{4}{c|||}{$\mathcal{C}_{11}$}&\multicolumn{4}{c||}{$\mathcal{C}_{12}$}\\
\cline{1-13}
(2,2)&144&0&0&0&264&0&0&0&576&0&0&0\\
\cline{1-13}
(3,2)&192&72&0&0&264&44&0&0&576&96&0&0\\
\cline{1-13}
(4,2)&312&0&0&0&308&0&0&0&672&0&0&0\\
\cline{1-13}
(5,2)&408&24&0&0&396&44&0&0&864&96&0&0\\
\cline{1-13}
(6,2)&648&144&72&0&572&88&44&0&1152&96&96&0\\
\cline{1-13}
(7,2)&1272&216&216&0&880&44&44&0&1536&0&0&0\\
\cline{1-13}
(8,1)&48&48&48&0&44&44&44&0&96&96&96&0\\
\cline{1-13}
(8,2)&2952&912&768&0&1584&308&176&0&2304&384&192&0\\
\cline{1-13}
Search w. (\ref{eqkd}) &\multicolumn{4}{c|||}{17 min.}&\multicolumn{4}{c|||}{8 min.}&\multicolumn{4}{c||}{4 min.}\\
\cline{1-13}
Search w. (\ref{eqif}) &\multicolumn{4}{c|||}{5 min.}&\multicolumn{4}{c|||}{2 min.}&\multicolumn{4}{c||}{2 min.}\\
\cline{1-13}
\end{tabular}
\end{table*}

\begin{table}[]
\centering
\caption{Multiplicities of  $(a,b)$ ETSs, LETSs, EASs and FEASs of Codes  $\mathcal{C}_{13}$, $\mathcal{C}_{14}$ and $\mathcal{C}_{15}$ within the range of $a \leq 10$ and $b \leq 2$}
\setlength{\tabcolsep}{.4pt}
\label{tab:50}
\begin{tabular}{||c|c|c|c|c|c|||c|c|c|c|c|||c|c|c|c|| } 
\cline{1-15}
&\multicolumn{5}{c|||}{$\mathcal{C}_{13}$} &\multicolumn{5}{c|||}{$\mathcal{C}_{14}$}&\multicolumn{4}{c||}{$\mathcal{C}_{15}$}\\
\cline{1-15}
$(a,b)$&Total&Total&Total&Total&Total&Total&Total&Total
&Total&Total&Total&Total&Total&Total\\
class&ETS&LETS&EAS&FEAS&ETS\cite{Fals}&ETS&LETS&EAS&FEAS&ETS\cite{Fals}&ETS&LETS&EAS&FEAS\\
\cline{1-15}
(2,2)&240&0&0&0&240&440&0&0&0&440&960&0&0&0\\
\cline{1-15}
(3,2)&216&0&0&0&216&396&0&0&0&396&864&0&0&0\\
\cline{1-15}
(4,2)&192&0&0&0&192&352&0&0&0&352&768&0&0&0\\
\cline{1-15}
(5,2)&168&0&0&0&168&308&0&0&0&308&672&0&0&0\\
\cline{1-15}
(6,2)&216&72&72&0&216&352&88&88&0&352&672&96&96&0\\
\cline{1-15}
(7,2)&408&24&24&0&408&572&0&0&0&572&768&0&0&0\\
\cline{1-15}
(8,2)&624&24&24&0&624&792&0&0&0&792&864&0&0&0\\
\cline{1-15}
(9,2)&912&120&120&0&912&968&44&44&0&968&1152&192&192&0\\
\cline{1-15}
(10,1)&0&0&0&0&0&44&44&44&0&44&0&0&0&0\\
\cline{1-15}
(10,2)&1560&168&168&0&1560&1276&88&88&0&1276&1728&0&0&0\\
\cline{1-15}
Search w. (\ref{eqkd}) &\multicolumn{5}{c|||}{38 min.}&\multicolumn{5}{c|||}{18 min.}&\multicolumn{4}{c||}{9 min.}\\
\cline{1-15}
Search w. (\ref{eqif}) &\multicolumn{5}{c|||}{12 min.}&\multicolumn{5}{c|||}{7 min.}&\multicolumn{4}{c||}{3 min.}\\
\cline{1-15}

\end{tabular}
\end{table}

In Table \ref{tab:50wifi}, we have also reported the multiplicity of instances of ETSs and TSs obtained by the non-exhaustive search algorithms of \cite{mehdi2012} and \cite{abu2010trapping}, respectively. 
As can be seen, there are some cases in Table \ref{tab:50wifi}, where the multiplicity of ETS classes obtained
here differs from the multiplicity of ETS classes reported in \cite{mehdi2012} and TS classes reported in \cite{abu2010trapping}. 
These cases are boldfaced in the table.

\begin{table}[]
\centering
\caption{Multiplicities of  $(a,b)$ ETSs, LETSs, EASs and FEASs of Code  $\mathcal{C}_{16}$ within the range of $a \leq 12$ and $b \leq 2$}
\setlength{\tabcolsep}{.5pt}
\label{tab:50wifi}
\begin{tabular}{||c|c|c|c|c|c|c|| } 
\cline{1-7}
&\multicolumn{6}{c||}{$\mathcal{C}_{16}$}\\
\cline{1-7}
$(a,b)$&Total&Total&Total&Total&Total&Total\\
class&ETS&LETS&EAS&FEAS&ETS\cite{mehdi2012}&TS\cite{abu2010trapping}\\
\cline{1-7}
(2,2)&810&0&0&0&810&nr\\
\cline{1-7}
(3,2)&729&0&0&0&729&nr\\
\cline{1-7}
(4,2)&648&0&0&0&648&648\\
\cline{1-7}
(5,2)&567&0&0&0&567&567\\
\cline{1-7}
(6,2)&486&0&0&0&486&486\\
\cline{1-7}
(7,2)&486&81&81&0&486&\textbf{485}\\
\cline{1-7}
(8,2)&648&81&81&0&648&\textbf{637}\\
\cline{1-7}
(9,2)&972&0&0&0&972&nr\\
\cline{1-7}
(10,2)&1377&81&81&0&1377&\textbf{1210}\\
\cline{1-7}
(11,2)&2106&324&324&0&\textbf{1944}&\textbf{1635}\\
\cline{1-7}
(12,1)&81&81&81&0&81&81\\
\cline{1-7}
(12,2)&3564&324&324&0&\textbf{2754}&\textbf{2166}\\
\cline{1-7}
Search w. (\ref{eqkd}) &\multicolumn{6}{c||}{2 min.}\\
\cline{1-7}
Search w. (\ref{eqif}) &\multicolumn{6}{c||}{1 min.}\\
\cline{1-7}
\end{tabular}
\end{table}

\begin{table*}[]
\centering
\caption{Multiplicities of $(a,b)$ ETSs, LETSs, EASs and FEASs of Codes  $\mathcal{C}_{17}$ and $\mathcal{C}_{18}$ within the range of $a \leq 10$ and $b \leq 2$}
\setlength{\tabcolsep}{1pt}
\label{tab:peg1}
\begin{tabular}{||c|c|c|c|c|c|c|c|c|||c|c|c|c|c|c|| } 
\cline{1-15}
&\multicolumn{8}{c|||}{$\mathcal{C}_{17}$} &\multicolumn{6}{c||}{$\mathcal{C}_{18}$}\\
\cline{1-15}
$(a,b)$ &Total&Total&Total&Total&Total&Total&Total&Total&Total&Total&Total &Total&Total&Total\\
class&ETS&LETS&EAS&FEAS&ETS\cite{Fals}&FEAS*&FEAS\cite{mehdi2012}&FAS\cite{kyung2012finding}&ETS&LETS&EAS&FEAS&FEAS*&FAS\cite{kyung2012finding}\\
\cline{1-15}
(2,2)&230&0&0&0&230&230&nr&230&916&0&0&0&458&458\\
\cline{1-15}
(3,2)&219&0&0&0&219&219&219&219&439&0&0&0&439&439\\
\cline{1-15}
(4,2)&208&0&0&0&208&208&208&208&420&0&0&0&420&420\\
\cline{1-15}
(5,2)&198&0&0&0&198&198&198&198&404&0&0&0&404&404\\
\cline{1-15}
(6,2)&207&19&0&0&207&205&205&205&388&0&0&0&387&387\\
\cline{1-15}
(7,1)&2&2&2&0&2&2&2&2&1&1&1&0&1&1\\
\cline{1-15}
(7,2)&276&24&24&0&276&271&271&271&406&30&30&0&403&403\\
\cline{1-15}
(8,1)&8&4&4&0&8&8&8&8&4&2&2&0&4&4\\
\cline{1-15}
(8,2)&466&61&60&0&466&458&458&458&524&45&44&0&519&519\\
\cline{1-15}
(9,1)&16&4&4&0&16&16&16&16&8&2&2&0&8&8\\
\cline{1-15}
(9,2)&870&75&74&0&870&855&855&nr&806&52&50&0&795&nr\\
\cline{1-15}
(10,1)&22&3&3&0&22&22&22&22&14&4&4&0&14&14\\
\cline{1-15}
(10,2)&1640&168&167&0&1640&1593&\textbf{1533}&nr&1305&73&73&0&1290&nr\\
\cline{1-15}
Search w. (\ref{eqkd}) &\multicolumn{8}{c|||}{20 min.}&\multicolumn{6}{c||}{21 min.}\\
\cline{1-15}
Search w. (\ref{eqif})&\multicolumn{8}{c|||}{11 min.}&\multicolumn{6}{c||}{11 min.}\\
\cline{1-15}

\end{tabular}
\end{table*}
In \cite{mehdi2012} and \cite{kyung2012finding}, for irregular codes, the authors relaxed the condition that degree-2 variable nodes of (fully)
absorbing sets must be connected to two satisfied check nodes. To compare our results for Codes $C_{17}$ and $C_{18}$ with those reported in \cite{mehdi2012} and \cite{kyung2012finding}, 
we have also reported the list of FEASs of these two codes with this modification in the definition of absorbing sets and fully absorbing sets. These results are identified with a star in Table \ref{tab:peg1}.
As can be seen, for both codes, these results match those obtained by the exhaustive search algorithm of~\cite{kyung2012finding}, in all the classes where the results are reported in ~\cite{kyung2012finding}.
Comparison with the non-exhaustive results of~\cite{mehdi2012}, however, shows a discrepancy in the multiplicity for the $(10,2)$ class.

We have also compared our results for Codes $C_{13}$, $C_{14}$, and $C_{17}$, with those obtained in~\cite{Fals}, in Tables~\ref{tab:50} and~\ref{tab:peg1}, respectively.
As can be seen, the multiplicities of ETSs for different classes match perfectly for all three codes with those reported in~\cite{Fals}. In terms of run-time, however, the proposed algorithm here is expectedly faster, particularly for $C_{14}$ and $C_{17}$. The run-times reported in~\cite{Fals} for the three codes are about $41$, $189$ and $273$ minutes, respectively.\footnote{The algorithm of~\cite{Fals} has been implemented in C++, and run on an Intel Core i7-2670QM 2.20 GHz laptop with 4 GB of RAM.} Notable here is that, while for our algorithm, the run-time for $C_{14}$ is less than that of $C_{13}$, the trend for the algorithm of \cite{Fals} is the opposite. In fact, unlike the brute force algorithms of \cite{Fals} and \cite{kyung2012finding}, where the complexity, in general, increases rapidly with the block length, the complexity of our algorithm, in general, decreases with the increase in the block length, for a fixed degree distribution. This can be seen by comparing the run-times for Codes $C_7$, $C_8$, and $C_9$, or Codes $C_{10}$, $C_{11}$, and $C_{12}$, or 
Codes $C_{13}$, $C_{14}$, and $C_{15}$. The reason for this behavior can be explained by the fact that the multiplicity of simple cycles of different length, which are the inputs to our search algorithm, is rather independent of the block length, and that as the block length increases the multiplicity of LETSs in many classes, particularly, those with larger $a$ and $b$ values decreases. 

\begin{figure}[] 
\centering
\includegraphics [width=0.6\textwidth]{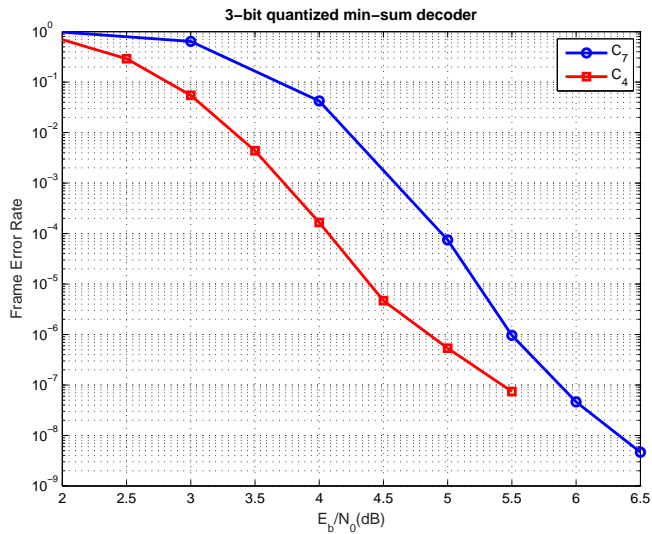}
\caption{FER results of Codes $\mathcal{C}_{4}$ and $\mathcal{C}_{7}$.}
\label{fig:FER}
\end{figure}

Finally, to demonstrate that the ETSs, discussed in this work, are in fact the main culprits in the error floor region of irregular LDPC codes, we 
perform Monte Carlo simulations to obtain the frame error rate (FER) of Codes $\mathcal{C}_{4}$ and $\mathcal{C}_{7}$, down to the start of their error floor region. 
For simulations, we consider binary phase shift keying (BPSK) modulation over an additive white Gaussian noise (AWGN) channel with coherent detection and a 3-bit quantized min-sum decoder. For each simulation point, we obtain 100 block errors. The FER results are presented in Fig. \ref{fig:FER}. 

For Code $\mathcal{C}_{4}$, at signal-to-noise ratio (SNR) of $5.5$ dB, all the 100 block errors correspond to ETSs, where 88 of those ETSs are within the range of Table~\ref{tab:34}. We expect the proportion of ETSs within the range of this table to increase by increasing the SNR.
The breakdown of the 88 ETSs is as follows: 28$\times(6,2)$, 26$\times(7,2)$, 16$\times(8,2)$, 9$\times(9,1)$, 6$\times(5,3)$, 1$\times(8,3)$, 1$\times(6,4)$ and 1$\times(7,4)$. 

For Code $\mathcal{C}_{7}$, at SNR of $6.5$ dB, among the 100 errors, 96 of them are ETSs, and out of this 96, 66 of them are ETSs within the range of Table~\ref{tab:34}. We expect that by increasing the SNR, both the proportion of  ETSs to the total errors, as well as the proportion of ETSs within the range of Table VII to total ETSs to increase.
From 66 ETSs, 63 are LETSs
 and 55 are LETSs in the $(7,1)$ class. 
The breakdown of the 66 ETSs is as follows: 55$\times(7,1)$, 8$\times(8,1)$, 2$\times(7,2)$ and 1$\times(8,2)$.

\section{Conclusion}
\label{sec:conclude}
In this paper, we proposed a hierarchical graph-based approach in the space of normal graphs to characterize elementary
trapping sets (ETSs) of irregular low-density parity-check (LDPC) codes. Two characterizations were proposed both based on three simple expansion techniques, called 
depth-one tree ($dot$), $path$ and $lollipop$, thus, the terminology {\em dpl characterization}. The proposed dpl characterizations, both, describe an ETS as an embedded sequence of ETS structures that starts from a simple cycle or a single variable node, and is expanded step by step through a combination of the three expansions to reach the ETS under consideration.  Corresponding to the first characterization, we demonstrated that the dpl characterization of $(a,b)$ leafless ETS (LETS) structures of 
variable-regular Tanner graphs with a properly selected variable degree $d_\mathrm{v}$, over a properly chosen range $a \leq a'_{max}$ 
and $b \leq b'_{max}$, can be used to exhaustively cover all the normal graphs of all the non-isomorphic $(a,b)$ LETS structures 
of irregular Tanner graphs with a given variable node degree distribution $\lambda(x)$, over any desired range of $a \leq a_{max}$ and $b \leq b_{max}$,
where $d_\mathrm{v}$, $a'_{max}$, and $b'_{max}$, were derived as functions of $\lambda(x)$, $a_{max}$, and $b_{max}$.  
This characterization corresponds to an efficient exhaustive search algorithm for irregular LDPC codes with relatively small variable degrees, where one is interested in a rather small values of $a_{max}$ and $b_{max}$. For other scenarios, where the first characterization appeared inefficient, we developed the second dpl characterization 
of LETS structures of irregular Tanner graphs. This characterization is based on the application of all the possible $dot$, $path$ and $lollipop$ expansions to simple cycles of the graph, recursively. The characterization, and the efficiency of the corresponding exhaustive search algorithm, rely on a sequence of upper bounds $b^a_{max}$, on the $b$ values for the classes of $(a,b)$ LETS structures that will need to be covered in the process, for the values of $a$ in the range $g/2 \leq a \leq a_{max}$. Such upper bounds were derived using a backward recursion with the initial condition that $b^{a_{max}}_{max} = b_{max}$. 

In summary, the proposed characterizations/search algorithms can be considered as the generalization of our earlier work~\cite{hashemireg} on LETSs of variable-regular LDPC codes to the cases where the code has variable nodes with a variety of degrees, and to ETSs that are not leafless. 
To the best of our knowledge, the proposed graph-based search algorithm is the most efficient exhaustive search algorithm available for finding ETSs of irregular LDPC codes.
It is also the most general, in that, it is applicable to codes with any degree distribution, girth, rate and block length. In particular, compared to the brute force exhaustive search algorithms of~\cite{kyung2012finding} and~\cite{Fals}, that are limited to short to moderate block lengths, our dpl-based search algorithm has no such limitation.

\end{document}